\documentclass[twocolumn,prl,aps,floatfix,longbibliography,superscriptaddress]{revtex4-2}

\usepackage{color}
\usepackage{graphicx}
\usepackage{siunitx}
\usepackage[color=green!60]{todonotes}
\usepackage{physics}
\usepackage{amsthm}
\usepackage{amsmath}
\usepackage{amssymb}
\usepackage{enumerate}
\usepackage{placeins}
\usepackage{mathtools}
\usepackage{booktabs}
\usepackage{dsfont}
\usepackage{yfonts}
\usepackage{calligra}
\usepackage{hyperref}
\usepackage{svrsymbols} 
\usepackage{tikz}
\usepackage{multibib}
\usepackage[version=4]{mhchem}
\usepackage{relsize}  

\usepackage{pdfpages} 
\usepackage{pgffor} 

\DeclareMathAlphabet{\mathcalligra}{T1}{calligra}{m}{n}

\DeclareSIUnit\angstrom{\protect\text{Å}}

\newcommand{\he}{\ce{^4He}}

\newcommand{\be}{\begin{equation}}
\newcommand{\ee}{\end{equation}}

\newcommand{\bsube}{\begin{subequations}}
\newcommand{\esube}{\end{subequations}}
\newcommand{\ba}{\begin{array}}
\newcommand{\ea}{\end{array}}

\newcommand{\bea}{\begin{eqnarray}}
\newcommand{\eea}{\end{eqnarray}}

\newcommand{\bfr}{\mathbf{r}}
\newcommand{\vrg}{\!\mathbf{\mathlarger{\mathlarger{\mathcalligra{r}}}}}

\AtBeginDocument{%
\heavyrulewidth=.08em
\lightrulewidth=.05em
\cmidrulewidth=.03em
\belowrulesep=.65ex
\belowbottomsep=0pt
\aboverulesep=.4ex
\abovetopsep=0pt
\cmidrulesep=\doublerulesep
\cmidrulekern=.5em
\defaultaddspace=.5em
}

\usepackage{pdfpages} 
\usepackage{pgffor} 
\usepackage{xr} 

\makeatletter
\AtBeginDocument{\let\LS@rot\@undefined}
\makeatother

\def\supplementfilename{supplement}

\pdfximage{\supplementfilename.pdf}
\def\numbersupplementpages{\the\pdflastximagepages}

\newif\ifarXiv
\arXivtrue 
\usepackage{hyperref}
\hypersetup{
    colorlinks=true,
    linkcolor=blue,
    filecolor=magenta,      
    urlcolor=blue,
    citecolor=blue,
}

\begin{document}

\title{Strain-induced superfluid transition for atoms on graphene}

\author{Sang Wook Kim}
\affiliation{Department of Physics, University of  Vermont, Burlington, VT 05405, USA}
\author{Mohamed Elsayed}
\affiliation{Department of Physics, University of  Vermont, Burlington, VT 05405, USA}
\author{Nathan S. Nichols}
\affiliation{Data Science and Learning Division, Argonne National Laboratory, Argonne, Illinois 60439, USA}
\author{Taras Lakoba} 
\affiliation{Department of Mathematics \& Statistics, University of  Vermont, Burlington, VT 05405, USA}
\author{Juan Vanegas}
\affiliation{Department of Physics, University of Vermont, Burlington, VT 05405, USA}
\author{Carlos Wexler} 
\affiliation{Department of Physics and Astronomy, University of Missouri, 
Columbia, MO 65211, USA}
\author{Valeri N.  Kotov}
\affiliation{Department of Physics, University of  Vermont, Burlington, VT 05405, USA}
\author{Adrian Del Maestro}
\affiliation{Department of Physics and Astronomy, University of Tennessee, Knoxville, TN 37996, USA}
\affiliation{Min H.~Kao Department of Electrical Engineering and Computer Science, University of Tennessee, Knoxville, TN 37996, USA}

\date{\today}

\begin{abstract}
Bosonic atoms deposited on atomically thin substrates represent a playground for exotic quantum many-body physics due to the  highly-tunable, atomic-scale nature of the interaction potentials.   The ability to engineer strong interparticle interactions can lead to the emergence of complex collective atomic states of matter, not possible in the context of dilute atomic gases confined in optical lattices. While it is known that the first layer of adsorbed helium on graphene is permanently locked into a solid phase, we show by a combination of quantum Monte Carlo and mean-field techniques, that  simple isotropic graphene lattice expansion effectively unlocks a large variety of two-dimensional ordered commensurate, incommensurate, cluster atomic solid, and superfluid states for adsorbed atoms.  It is especially significant that an atomically thin superfluid phase of matter emerges under experimentally feasible strain values, with potentially supersolid phases in close proximity on the phase diagram. 
\end{abstract}

\maketitle

\section{Introduction}

The quest to understand the behavior of strongly interacting electrons in quantum materials 
has led to a fruitful program of quantum simulation \cite{Altman:2021qm}, where
analogous quantum systems are constructed from well-understood and controllable
constituents.  Promising examples include the study of atoms confined in
optical lattice potentials \cite{Bloch:2008,Lewenstein,Zhang2018}, electrons in
two-dimensional (2D) materials -- most notably graphene and its derivatives --
\cite{Antonio, vdwgeim}, Rydberg arrays \cite{Semeghini:2021dl}, and
superconducting quantum circuits \cite{Ma:2019zp}. Many of these approaches can
realize lattice Hamiltonians on mesoscopic scales, or at low densities;
however, generating strong interactions at the atomic scale remains a
challenge.  A promising route is the construction of synthetic matter where atoms are adsorbed onto a physical substrate solid with \cite{Ming:2022zm} or without chemical bonding \cite{Kreisel:2021xx,DelMaestro:2021kc}.

\begin{figure*}[t]
\centering
  \includegraphics[width=1.9\columnwidth]{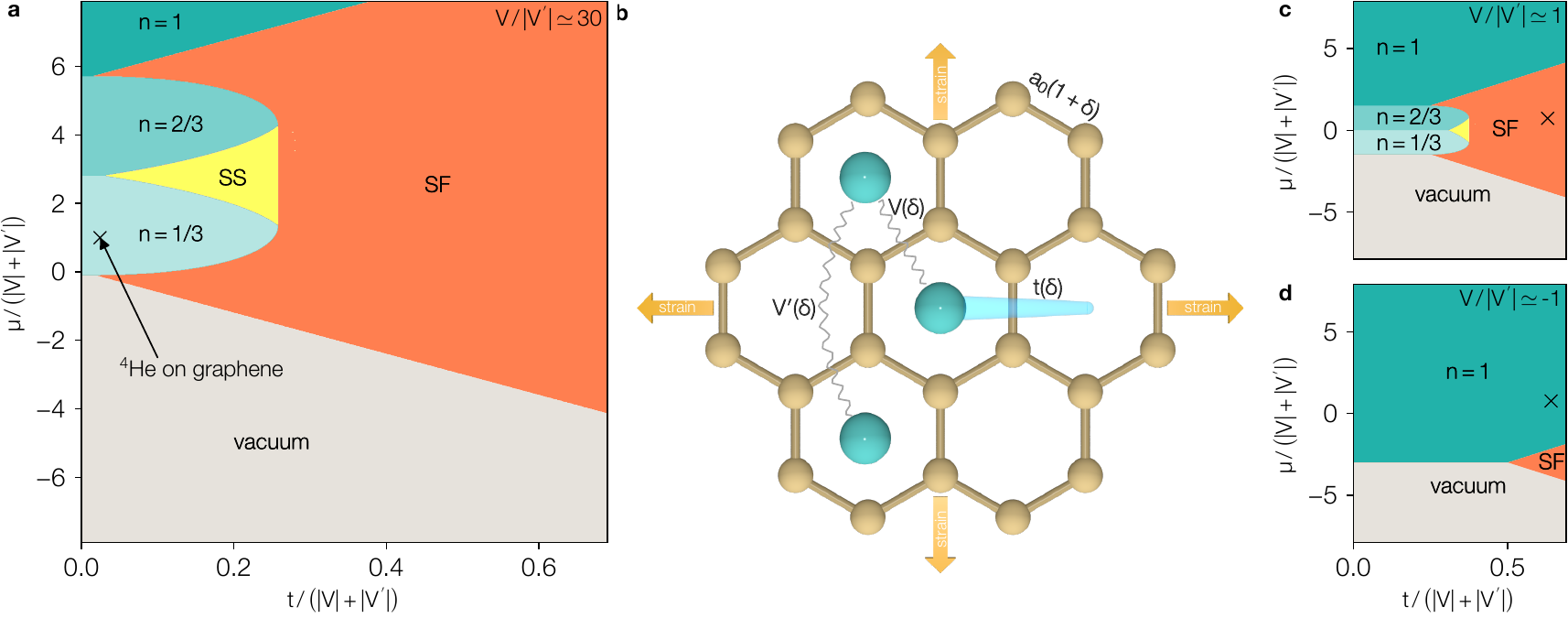}
\caption{\textbf{Strain-tuning the phase diagram.} 
\textbf{a} The mean field phase diagram as a function of dimensionless chemical potential $\mu$ and hopping $t$ for hard-core bosons on the 2D triangular lattice with the ratio of nearest to next-nearest neighbor interactions $V/\abs{V'} \simeq 30$ (the physically realized value as indicated by a $\times$ for \he{} on graphene \cite{Yu:2021tw} at fixed small $\mu$).  Lobes with crystalline insulating phases appear at filling fractions $n=1/3,\,2/3,$ and $1$ as $\mu$ is increased for small $t$, and a superfluid phase (SF) is stable for larger $t$.  A supersolid (SS) can exist between the lobes. Realizing non-solid order would require pushing the \he{} on graphene system to larger values of $t/(\abs{V} + \abs{V^\prime})$; however, the only experimental tuning parameter ($\mu$) is restricted to vertical movement in the phase diagram.   \textbf{b} The effective Bose--Hubbard Hamiltonian describing \he{} atoms adsorbed on a suspended graphene membrane can be mechanically manipulated via isotropic biaxial strain that increases the C--C bond length by a fraction $\delta$. The resulting microscopic hopping $t$, nearest-neighbor $V$, and next-nearest neighbor $V^\prime$ interactions are found to be strong functions of $\delta$ (see Fig.~\ref{fig:VGtVV}), opening up the possibility of realizing superfluid order. \textbf{c} and \textbf{d} are modified mean field phase diagrams in the presence of strain for $V/\abs{V^\prime} \simeq 1$ and $V/\abs{V^\prime} \simeq -1$, respectively. The large reduction and change in sign of the nearest neighbor interaction $V$ as a function of $\delta$ has a drastic effect: \he{} on strained graphene (indicated by $\times$) may become superfluid, or admit a strongly correlated $n=1$ insulator.}
\label{fig:MF}
\end{figure*}

Here we consider the latter case and study bosonic \he{} atoms adsorbed on graphene, where both atom-atom and atom-substrate interactions are driven by Van der Waals (VdW) dispersion forces.  This is an ideal platform to study fundamental many-body phenomena such as the formation of solid, superfluid, and supersolid phases, as well as the quantum phase transitions between them.  Adsorbed \he{} films have been a subject of considerable interest for over half a century and have drastically informed our understanding of criticality, including the role of the healing length \cite{Henkel:1969gx} and the universal jump of the superfluid density at the Kosterlitz-Thouless transition \cite{Agnolet:1989ou}. For a flat crystalline substrate such as graphite, the presence of strong adsorption sites forming a triangular lattice (the dual lattice corresponding to graphite hexagon centers) produces a series of commensurate and incommensurate solid phases in the first layer of adsorbed \he{} observable by anomalies in the heat capacity \cite{Bretz:1971jo,Bretz:1973ky,Zimmerli:1988ii, Greywall:1991ns}.  In the second and further adsorbed layers, the interacting bosonic $^4$He atoms can form a superfluid phase at a temperature $T_{\rm KT}$ below the bulk $T_\lambda$, which can be detected via third sound or a frequency shift of the adsorbed mass with torsional oscillator measurements \cite{Zimmerli:1992hz,Crowell:1996kn,Nyeki:1998fk}.  Further details on the structure of the adsorbed phases and the resulting coverage--temperature phase diagram have been obtained by extensive numerical simulations exploiting various levels of approximation for the graphite--helium interaction \cite{Whitlock:1998gb,Corboz2008cb,Pierce:2000cj,Ahn:2016dm}. While it has ultimately been understood that bulk helium does not exhibit a supersolid phase -- one that simultaneously breaks translational and gauge symmetries -- the existence of supersolidity in models of hard-core bosons on the triangular lattice \cite{Wessel:2005ik} makes adsorbed helium on graphite a potential platform for realizing exotic phases.  Recent experimental results provide support for this scenario, arguing for intertwined superfluid and density wave order \cite{Nakamura:2016wf,Nyeki:2017ef,Choi:2021vy} for multiple layer helium films on pristine graphite surfaces. However, the first layer remains strongly bound to the surface and displays no evidence of superfluid behavior.

The propensity for insulating behavior of the helium atoms close to the substrate is a result of the relatively strong corrugation potential ($\sim \SI{30}{\kelvin}$ from peak to valley), that localizes atoms through an exponential suppression of tunneling between triangular lattice adsorption sites.  It is thus natural to consider replacing the graphite substrate with graphene, providing the same triangular lattice of adsorption sites, but with an attractive potential approximately 10\% weaker.  This has motivated a number of theoretical studies employing ab initio quantum Monte Carlo simulations \cite{Gordillo:2009jb,Gordillo:2011jb,Gordillo:2012fl,Kwon:2012ie,Happacher:2013ht,Gordillo:2014cp,Markic:2016dm}; however, the first layer appears to remain stubbornly insulating, providing little motivation for expanded experimental searches.

In this manuscript, we propose that quantum delocalized atomically thin superfluid phases of \he{} can be realized in this system through the application of even moderate (5--15\%) biaxial (isotropic) strain to the graphene membrane. 
This is possible due to the fact that graphene --  an atomically thin solid itself -- can be mechanically strained along one \cite{Hone-PNAS,Naumis2017},  or multiple axes \cite{Zabel2012,Androulidakis2015} to produce an isotropic increase in the carbon--carbon bond length.  The extreme sensitivity of the system to strain arises due to the fact that the interaction between adsorbed \he{} atoms changes from a strong (hard-core) repulsion to a weak attractive VdW tail on the Angstrom scale \cite{Przybytek:2010ol}, which is also the scale of the underlying graphene lattice potential. Thus small changes in the latter can lead to a very strong modification of atomic interactions. Consequently, graphene's lattice potential can be viewed as an ``effective 2D lattice" for the \he{} atoms with a period on the scale of atomic interactions. This setup is conceptually impossible to achieve for conventional dilute gases in optical lattices \cite{Bloch:2008} which are soft-core, allowing multiple bosons per site, and may have only tunable kinetic energy. Instead, \he{} on graphene can realize an effective 2D hard-core Bose--Hubbard model with strain-dependent nearest ($V$) and next-nearest ($V'$) neighbor interactions.  Our intuitive picture is motivated by mean field calculations (Figure~\ref{fig:MF}) and confirmed with large scale \emph{ab initio} quantum Monte Carlo simulations of helium on strained graphene at low temperature that are finite size scaled to the thermodynamic limit.  We conclude that this system is a highly tunable (via mechanical strain and pressure/chemical potential) platform for the experimental exploration and discovery of strictly two dimensional strongly interacting quantum phases of matter. 

\section{Characterization and Strain-Tuning}

\he{} atoms of mass $m_4$ interacting with a biaxially strained suspended graphene membrane can be described by the microscopic many-body Hamiltonian: 
\begin{equation}
    H = -\frac{\hbar^2}{2m_4} \sum_{i=1}^N \vb*{\nabla}_i^2 + \sum_{i=1}^N \mathcal{V}_{\rm \graphene}(\vb*{r}_i;\delta) + 
\sum_{i <j} \mathcal{V}(\vb*{r}_i-\vb*{r}_j)\, .
\label{eq:Ham}
\end{equation}
Here, $\mathcal{V}_{\graphene}$ is the adsorption potential experienced by an atom at spatial position $\vb*{r}_i$ with strain captured by $\delta \equiv{a/a_0-1}$ quantifying the increase of the carbon--carbon distance $a$ with respect to its unstrained as-grown value $a_0 \simeq \SI{1.42}{\angstrom}$.  $\mathcal{V}_{\graphene}$ can be obtained by summing up all the individual VdW interactions between \he{} and the C atoms in the membrane, carefully considering the effects of strain on the electronic polarization of graphene itself \cite{Nichols:2016hd} (see Methods section for more details). The interaction between \he{} atoms is captured by $\mathcal{V}$ which is known to high precision \cite{Przybytek:2010ol,Cencek:2012iz}. Numerical simulations of Eq.~\eqref{eq:Ham} with $\delta = 0$ at low temperature are consistent with the experimentally observed phase diagram for graphite. They demonstrate that as the pressure is increased from vacuum, there is a first order transition where a single layer is adsorbed, forming a commensurate incompressible solid phase dubbed C1/3 where \he{} atoms are localized around 1/3 of the strong binding sites of the \SI{30}{\degree} rotated triangular lattice with lattice constant $\sqrt{3}a_0$ corresponding to graphene hexagon centers.  The C1/3 phase is stable over a range of chemical potentials \cite{Zimanyi:1994rk, Happacher:2013ht,Yu:2021tw} due to the strong repulsive interactions that induce an energy cost of $\mathcal{V}(\sqrt{3}a_0) \approx \SI{50}{\kelvin}$ per atom when nearest neighbor triangular sites are occupied increasing the filling beyond $n=1/3$.  As the pressure of the proximate helium gas is further increased, eventually other commensurate and incommensurate phases can be realized due to energetic compensation by the chemical potential, including those with proliferated domain walls \cite{Happacher:2013ht}. Beyond a triangular lattice filling fraction of $n \simeq 0.6$, it is energetically favorable to form a second layer (and beyond), but at all lower fillings, the width of the transverse wavefunction of the adsorbed atoms remains on the atomic scale (see supplemental Fig.~3).

This strongly 2D character was recently exploited to demonstrate that the first adsorbed layer of helium on unstrained graphene ($\delta = 0$) is well characterized by an effective extended hard-core 2D Bose--Hubbard model \cite{Yu:2021tw,DelMaestro:2021kc} with hopping $t$ and both nearest ($V$) and next-nearest neighbor ($V^\prime$) density--density interactions on the triangular lattice:  
\begin{align}
{H}_{BH} &=-t\sum_{\left\langle i,j\right\rangle }\left({b}_{i}^{\dagger}{b}_{j}^{\phantom \dagger}+{b}_{j}^{\dagger}{b}_{i}^{\phantom \dagger}\right)+V\sum_{\left\langle i,j\right\rangle }{n}_{i}{n}_{j}
\nonumber \\
         &\quad +V^{\prime}\sum_{\left\langle \left\langle i,j\right\rangle \right\rangle }{n}_{i}{n}_{j}-\mu\sum_{i}{n}_{i}\, .
\label{eq:BH}
\end{align}
Here $b^{\dagger}_i(b^{\phantom \dagger}_i)$ creates(annihilates) a hard-core \he{} atom on site $i$ of the triangular lattice and $n_i = 
b^{\dagger}_ib^{\phantom \dagger}_i$ measures the number of atoms per site where $[b^{\phantom \dagger}_i,b^{\dagger}_j] = \delta_{ij}$. $\expval{i,j}$ and $\expval{\expval{i,j}}$ indicate nearest and next-nearest neighbors respectively.

For \he{} on unstrained graphene ($\delta = 0$) it is known from many-body as well as first principle ab initio methods \cite{Yu:2021tw} that $V$ is strongly repulsive, originating from the overlap of localized wavefunctions on the scale of the lattice spacing, while 
$V^\prime$ is much weaker and attractive, due to the VdW tail with the ratio $V/\abs{V^\prime} \simeq 30$.  

\begin{figure}
\includegraphics[width=\columnwidth]{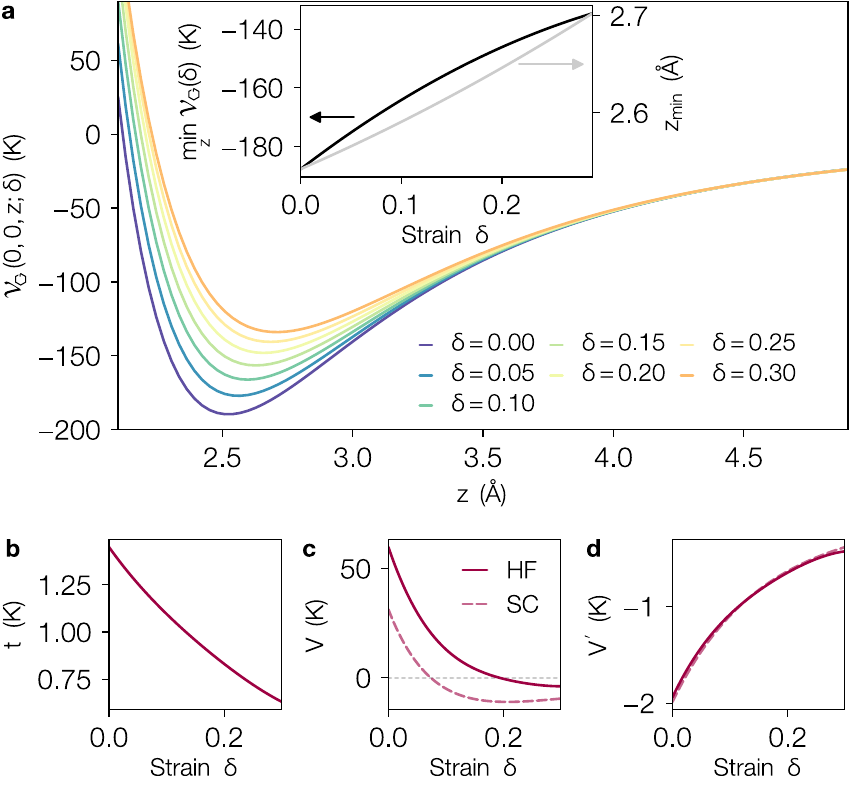}
\caption{\textbf{Strain-dependent adsorption and model parameters.} \textbf{a} The microscopic helium--graphene adsorption potential $\mathcal{V}_{\graphene}$ that appears in Eq.~\eqref{eq:Ham} as a function of the height of an atom situated directly above a strong graphene adsorption site for different values of strain $\delta$.  The inset quantifies how strain leads to a softening of the potential by reducing the binding energy per atom (left axis), as well as moving the location of the minimum, $z_{\rm min}$ (right axis) further from the membrane. Panels $\textbf{b} - \textbf{d}$ demonstrate the effects of biaxial strain on the effective 2D Bose--Hubbard model parameters $t$, $V$, and $V^\prime$ as computed via Hartree--Fock (HF) using the same method as described in Ref.~\cite{Yu:2021tw}.  Dashed lines correspond to the semi-classical (SC) predictions for the interaction parameters computed directly from the inter-atomic potential $\mathcal{V}$ evaluated at the expanded  nearest and next-nearest neighbor distances assuming $\delta$-function wavefunctions. Panel \textbf{c} highlights the fact that at large strain, the nearest neighbor interaction $V$ can vanish, before changing signs and becoming attractive, which can lead to a drastic modification of the unstrained phase diagram.}
\label{fig:VGtVV}
\end{figure}

The phase diagram of Eq.~\eqref{eq:BH} can be directly computed at the mean field level \cite{Murthy-phase-diag} (see Methods), and it exhibits insulating phases at commensurate filling fractions $n=1/3,\,2/3,\,1$, as well as a superfluid and supersolid phase as a function of the dimensionless chemical potential $\mu$ and hopping $t$ as shown in Fig.~\ref{fig:MF}\textbf{a}. The physical system of \he{} on graphene at fixed small $\mu$ is indicated by a cross ($\times$).  By tuning the chemical potential (through the pressure of the proximate \he{} gas) an experiment could in principle observe the first order adsorption transition to a $n=1/3$ solid in the first layer. However,  
as the carbon atoms are moved further apart via strain, we expect that both $V$ and $V^\prime$ should be changed (Fig.~\ref{fig:MF}\textbf{b}) leading to qualitative changes in the mean field phase diagram (as shown in Figure~\ref{fig:MF}\textbf{c} and \textbf{d}). This culminates in access to both a superfluid and strongly correlated fully filled insulating phase as $V/V^\prime$ is reduced through zero.   

To validate this simple picture and understand how the strain dependence of these effective parameters is generated in the physical helium on graphene system, we can analyze the adsorption potential $\mathcal{V}_{\graphene}$ and resulting effective parameters of the 2D Bose-Hubbard model for different values of $\delta$ as shown in Figure~\ref{fig:VGtVV}.

\begin{figure*}
\begin{center}
\includegraphics[width=\textwidth]{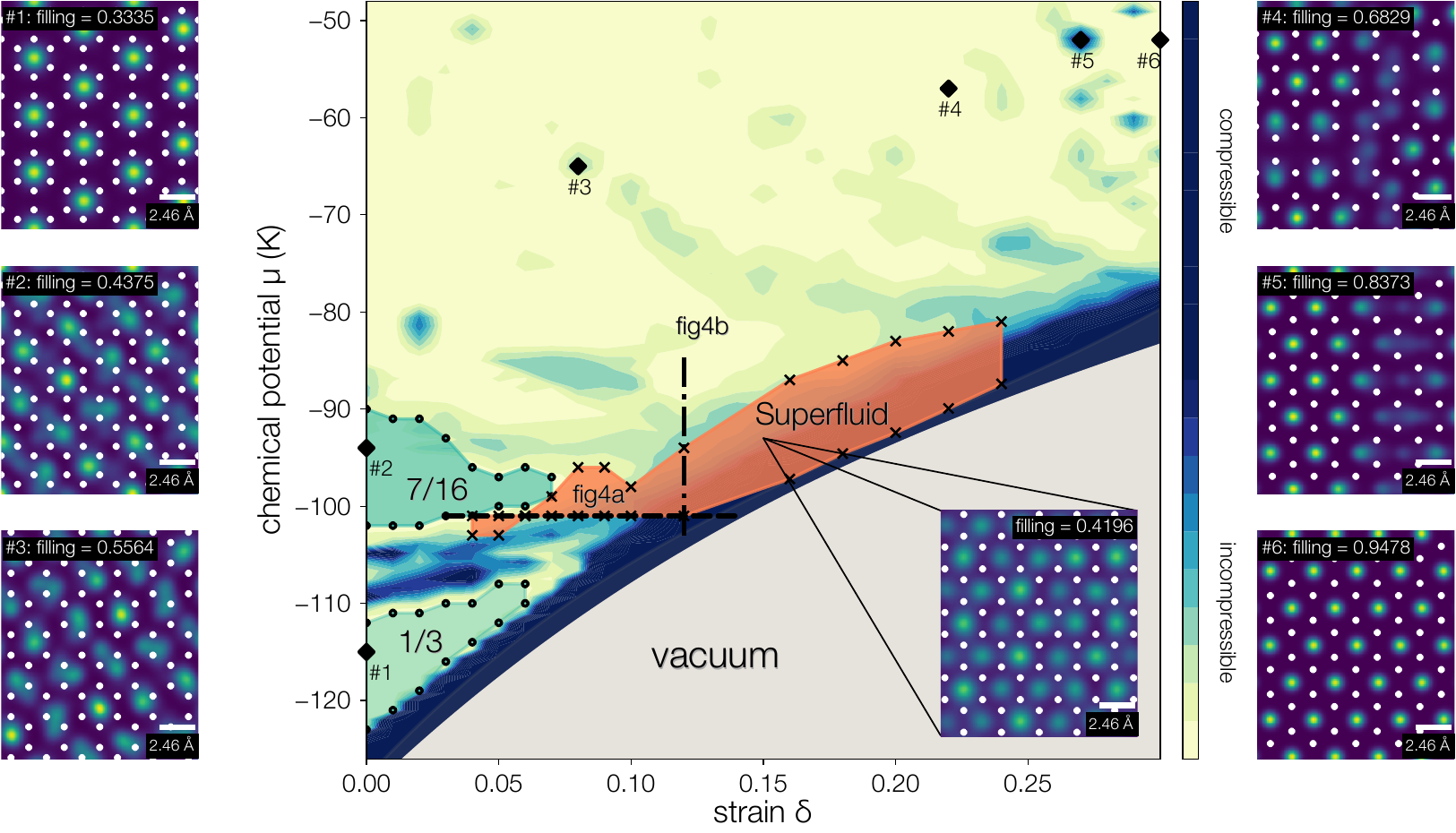}
\end{center}
\caption{\textbf{Superfluid phase diagram for helium adsorbed on strained graphene.} Quantum Monte Carlo adsorbed particle configurations (halos \#1--\#6) and phase diagram (central panel) as a function of chemical potential $\mu$ and isotropic strain $\delta = a/a_0 -1$ (enhancement of the carbon-carbon distance).  The colored background of the main panel shows the compressibility of the adsorbed layer, with darker colors being more compressible.  At small values of strain, large colored regions indicate commensurate solid phases with $n=1/3$ and $7/16$ of adsorption sites filled. As strain is increased, a superfluid phase (SF, orange) emerges with a critical temperature above $T = \SI{0.5}{\kelvin}$ in thermodynamic limit, with a boundary indicated by $\times$ symbols. The behavior of the superfluid and particle density along the indicated horizontal and vertical cuts is shown in Figure~\ref{fig:pdLower}. Sub-panels (halos \#1--\#6) show the average density of adsorbed helium at different regions of the phase diagram (indicated by $\blacklozenge$), detailing the particle configurations in both commensurate and incommensurate phases for $N_{\graphene}=48$ adsorption sites at $T=\SI{1.0}{\kelvin}$. The filling fraction is indicated in each halo along with a scale bar highlighting the increase in the C-C separation, and the colorscale is set such that $\int \dd{x}\dd{y} \rho(x,y) = N$.  At large strain, the unit filling phase predicted by mean field theory is observed in halo \#6. It is important to note the adsorbed density in the superfluid phase (bottom right of central panel) is qualitatively different, exhibiting the delocalization of atoms between all adsorption sites.
}
\label{fig:pimcPD}
\end{figure*}
In panel \textbf{a}, we demonstrate that for a single \he{} atom at position $z$ directly above the center of a graphene hexagon, as strain is increased, the adsorption potential becomes less attractive, with a minimum that softens by 30\% from $\SI{-188}{\kelvin}$ for $\delta = 0$ to $\SI{-134}{\kelvin}$ for $\delta = 0.3$ corresponding to extreme strain as quantified in the inset.  The location of the adsorbed 2D layer at this minimum, $z_{\rm min}$, is also pushed further from the sheet by 7\%, from \SI{2.5}{\angstrom} to \SI{2.7}{\angstrom} yielding a concomitant reduction in the effect of the corrugation potential.  This means, in essence, that  the increase of $z_{\rm min}$  is correlated with the decrease
of the effective barrier height related to the in-plane potential   (calculated at $z_{\rm min}$), in turn, for example, causing an increase of atomic delocalization as a function of strain.  These changes in the microscopic adsorption potential are reflected in the effective parameters of the Bose--Hubbard model as computed via Hartree--Fock calculations with the results shown in Figure~\ref{fig:VGtVV}\textbf{b}-\textbf{d} (see Methods section for details).  They are calculated from the average interaction energy at the nearest and next-nearest neighbor level determined from the self-consistent adsorbed wavefunctions and compared with the strain dependence of the semi-classical (SC) predictions $V_{\rm SC} \equiv \mathcal{V}(\sqrt{3}(1+\delta)a_0)$ and $V^\prime_{\rm SC} \equiv \mathcal{V}(3(1+\delta)a_0)$ computed directly from the \he{} --\he{} interaction potential.  Here, SC refers to the use of point-particle wavefunctions with the full quantum potential $\mathcal{V}$.
As can be seen in Figure~\ref{fig:VGtVV}\textbf{b}, the nearest neighbor interaction experiences a drastic reduction as $\delta$ is increased, with strong wavefunction renormalization effects, and vanishes near 19\% strain, before becoming attractive for larger strains. As $V^\prime$ is controlled by the tail of the long-distance VdW interactions, and wavefunction effects have already been built into $\mathcal{V}$ at this scale, there is nearly perfect agreement between the semi-classical and the Hartree--Fock calculation.


\section{Superfluid Phase Diagram}
The drastic decrease of $V$ in the effective Bose--Hubbard description due to strain engineering provides a route to increase the dimensionless hopping parameter $t/\qty(\abs{V(\delta)} + \abs{V^\prime(\delta)})$ that controls the transition to the superfluid phase as detailed by the movement of the cross in Fig.~\ref{fig:MF}.  However, a number of questions remain regarding whether or not this is a realistic scenario for helium on graphene, as the extended Bose--Hubbard model describes only its 2D low energy sector, with the microscopic system allowing for a plethora of phases not present in the lattice model \cite{Ahn:2016dm}.  To validate these predictions, and generate a physical phase diagram, we have performed \textit{ab initio} quantum Monte Carlo simulations of the full microscopic Hamiltonian in Eq.~\eqref{eq:Ham} for temperatures below $T_{\lambda} \simeq \SI{2.17}{\kelvin}$, and over a wide range of chemical potentials and isotropic biaxial strains, up to 30\%.  The details of our simulations, based on the Feynman path integral formalism, are included in the Methods section, along with a description of how finite size graphene simulations for cells with dimension $L_x \times L_y$ were combined to extrapolate to the thermodynamic limit.  

The combination of these simulations constitutes the most important result of this work, presented as a chemical potential $\mu$ -- strain $\delta$ phase diagram in Figure~\ref{fig:pimcPD}.  Here, the main panel shows the strain dependence of the first order phase transition from vacuum to a single adsorbed layer (see Supplementary Figure~4 for more details). It is pushed to larger chemical potentials as $\delta$ is increased, consistent with the softening of the adsorption potential highlighted in Figure~\ref{fig:VGtVV}. At low strain ($<5\%$), as the chemical potential is increased, we find commensurate phases with filling fractions $n=1/3$ and $n=7/16$ that have been previously observed in simulations of unstrained graphene \cite{Happacher:2013ht} as well as experiments on graphite \cite{Greywall:1991ns}. We note that the $n=7/16$ solid is not realized in the discrete lattice model, and is only energetically favorable in the presence of a continuous adsorption potential.  As strain is further increased, the strong nearest-neighbor repulsion is reduced as the triangular lattice adsorption sites are moved further apart and the adsorbed layer moves further away from the membrane. Above 5\% strain, there is a small range of fine-tuned chemical potentials near $\mu/k_{\rm B} = \SI{-101}{\kelvin}$ where there is a transition from either a low density compressible liquid or vacuum to a superfluid.   Here, superfluidity is quantified within the two-fluid picture where the total adsorbed density of atoms is broken into a normal and superfluid part with $\rho = \rho_n + \rho_s$ where $\rho = \expval{N}/(L_x L_y)$ with $N$ the average number of adsorbed \he{} atoms.  While this sliver persists in the thermodynamic limit, the superfluid density develops an aspect ratio dependence suggesting this region could be non-universal. For larger values with $\delta > 0.1$, the extended superfluid region is more robust, extending up to 25\% strain and over a range of chemical potentials. Within the superfluid phase, there is some evidence of competing solid order, but further work remains to be done to confirm the existence of a supersolid phase induced by the strained graphene lattice potential.

Surrounding the phase diagram are six halo figures showing the average density of particles superimposed on the strained graphene lattice, 
with the scale-bar representing the nearest neighbor distance for unstrained graphene $\sqrt{3}a_0 \simeq \SI{2.46}{\angstrom}$.  For $\delta = 0$, commensurate phases with $n=1/3$ and $n=7/16$ are shown, while at larger strain and higher filling, phases with domain walls (e.g.~\#4 with $n\simeq0.68$) are observed. For the strongest values of strain, the interaction between adsorbed \he{} atoms is dominated by the attractive long-range VdW tail, and a phase with unit filling is clearly observed as predicted by mean field theory (see Figure~\ref{fig:MF}).  Supplemental Figure~1 depicts the mean field phase diagram in the same physical units considered here. While there are quantitative modifications of boundaries, its major features are recovered with the exception of the $n=2/3$ insulator which is not energetically favorable in the presence of a continuum lattice potential. 

\begin{figure}[t]
\begin{center}
\includegraphics[width=0.99\columnwidth]{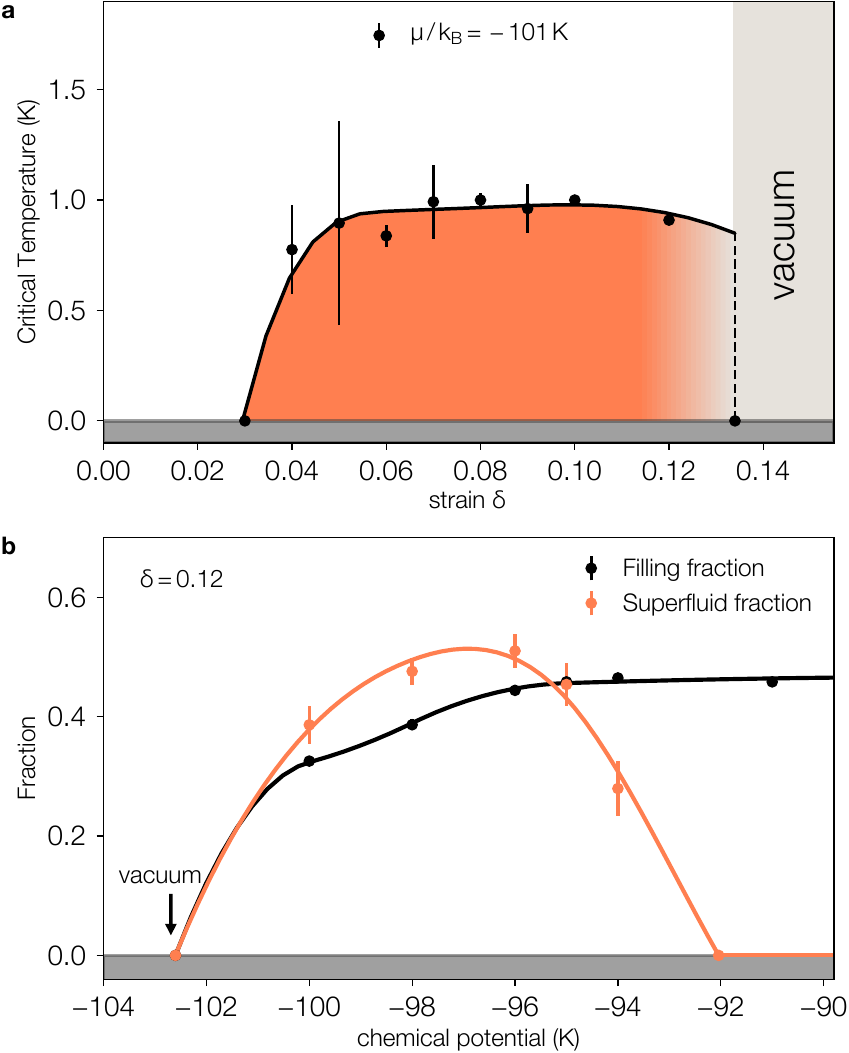}
\end{center}
\caption{\textbf{Details of the superfluid phase.} \textbf{a} A horizontal cut at fixed chemical potential $\mu/k_{\rm B} = \SI{-101}{\kelvin}$ indicated as a dashed line in Figure~\ref{fig:pimcPD} shows the finite temperature onset of the superfluid phase as a function of strain. A non-vanishing superfluid fraction is measured up to the transition to the vacuum beyond $\delta = 0.13$. \textbf{b} The particle filing fraction $n$ and superfluid fraction $\rho_s/\rho$ of the adsorbed atoms as a function of chemical potential $\mu$ at a fixed value of strain $(\delta = 0.12)$ and temperature $T = \SI{0.6}{\kelvin}$ (vertical dot-dashed line in Figure~\ref{fig:pimcPD}). In both panels, the error bars are computed as a combination of stochastic uncertainties from the Monte Carlo simulations and errors obtained from extrapolating to the thermodynamic limit. The smooth lines represent a guide to the eye.}
\label{fig:pdLower}
\end{figure}
More details on the superfluid phase can be obtained by taking horizontal and vertical cuts through Fig.~\ref{fig:pimcPD} as shown in Figure~\ref{fig:pdLower}. For fixed $\mu/k_{\rm B} = \SI{-101}{\kelvin}$, the finite temperature strain phase diagram (panel \textbf{a}) shows the onset of a finite superfluid fraction near $\delta = 0.03$ and the transition to vacuum for $\delta > 0.13$.  The critical temperature for the onset of superfluidity remains relatively constant near $T\approx\SI{1}{\kelvin}$ over the entire phase. At a larger value of strain, $\delta = 0.12$, Figure~\ref{fig:pdLower}\textbf{b} tracks both the triangular lattice filling fraction $n$ and the superfluid fraction $\rho_s/\rho$ as a function of chemical potential, with the maximal signal occurring for a filling fraction $n \simeq 0.4$.

\section{Discussion and Prospects for Measurement}

Having analyzed the many-body phase diagram and identified a superfluid phase, it is natural to ask if this setup could be realized in a real experiment. Graphene is generally expected  to withstand uniaxial  mechanical strain of  around 20\% or more \cite{Lee385,Naumis2017}, and several percent strain has already been realized \cite{Cao2020, Hone-PNAS, Geim-uniaxial}. On a fundamental level, it is important that strain can lead to substantial qualitative changes in electronic properties, which can be calculated with great theoretical precision, and consequently affect measurable physical  characteristics  \cite{Naumis2017,maria}. In principle, extreme uniaxial strain ($\sim25 \%$) can also lead to quantum phase transitions in the electronic structure of graphene itself. For example,   a  merger of Dirac cones  can occur via a topological Lifshitz transition at a critical strain value \cite{10.1103/physrevb.80.045401}, leading to the creation of an insulating state. The concept of  ``strain engineering," \emph{i.e.}\@ manipulation of properties as a function of strain, is applicable to other classes of 2D materials as well, such as members of the dichalcogenide family (MoSe$_2$, MoS$_2$, WSe$_2$, WS$_2$) \cite{Naumis2017,maria,Roldan_2015}.  

The combination of a strain-engineered 2D  substrate with a proximate quantum gas opens up a new class of phenomena based on 2D material band structure modification. For graphene, isotropic biaxial strain -- as considered here -- is the ``simplest" theoretical form of strain, and  has been analyzed theoretically and realized experimentally  \cite{Naumis2017,Androulidakis2015,Carrascoso2022,Zabel2012,Roldan_2015}. 
As strain is applied equally along the armchair and zig-zag directions, the carbon--carbon  lattice spacing changes isotropically (as captured by $\delta$). In this case, the graphene electronic dispersion remains isotropic (and electronically stable at any strain value the material can support), but the van der Waals forces that adatoms experience on top of graphene are substantially modified.  As we have shown, the modification of interactions between bosonic adatoms can lead new low-dimensional quantum phases and quantum transitions between them.  The competition between superfluid and correlated solid orders throughout the phase diagram opens up the possibility of realizing an adsorbed supersolid phase induced by the graphene adsorption potential with broken gauge and lattice symmetries. Our simulations on finite size graphene membranes host a large number of commensurate and incommensurate solid phases proximate to the identified superfluid, but the realization of a thermodynamically stable supersolid phase in the small region of phase space predicted by the mean field theory remains numerically elusive at this time.  

In the laboratory, by combining well-known techniques to realize suspended graphene \cite{Meyer:2007ls} with state-of-the art protocols that can simultaneously measure positional and superfluid responses of atoms adsorbed on flat surfaces \cite{Yamaguchi:2022qt,Usami:2022vs,Choi:2021vy}, the phase diagram in Fig.~\ref{fig:pimcPD} could be experimentally explored.  Of course, it should be noted that such experiments are usually not performed under the ``ideal"  theoretical conditions assumed in our numerical modeling and could involve bending, proximity effects, strain asymmetry, etc.  However, due to the ultra-rapid pace of technological developments in the field of 2D materials it is reasonable that such experiments are feasible in the near future. 

\newpage

\section{Methods}

\subsection{Mean Field Theory}

Starting from the effective low-energy Bose--Hubbard Hamiltonian in Eq.~(\ref{eq:BH}), the interaction terms can be decoupled for each lattice site $i$ within the standard mean field approach, leading to:
\begin{align}
    {H}_{MF,i} &=-6t\qty[\psi\qty({b}^{\phantom \dagger}_{i}+{b}_{i}^{\dagger})-\psi^{2}] \nonumber \\
                   &\quad +6\left(V+V^{\prime}\right)\left(\rho{n}_{i}-\frac{\rho^{2}}{2}\right)-\mu{n}_{i},
\label{eq:HMF}
\end{align}
where we have introduced the condensate density ${\displaystyle \langle b^{\phantom\dagger}_{i}\rangle=\langle b_{i}^{\dagger}\rangle=\psi}$ and the localized density ${\displaystyle \langle n_{i}\rangle=\rho.}$ For an insulating state, $\psi = 0$. Diagonalizing in the basis of localized Wannier states \cite{Yu:2021tw}  gives the ground state energy (per lattice site)
\begin{align}
    E &=6t\psi^{2}+3\left(V+V^{\prime}\right)\rho\left(1-\rho\right) - \frac{\mu}{2} \nonumber \\
      & \quad -\sqrt{\left(\frac{\mu-6\left(V+V^{\prime}\right)\rho}{2}\right)^{2}+\left(6t\psi\right)^{2}}\, .
\end{align}
The self-consistent eigenstates can be found
by solving ${\displaystyle \partial_{\rho}E=\partial_{\psi}E=0}$,
yielding the particle and condensate densities as:
\begin{align}
    \label{eq:rho}
\rho & =\frac{6t+\mu}{12t+6\left(V+V^{\prime}\right)}\\
\psi & =\frac{\sqrt{\left(6t+\mu\right)\left(6t+6\left(V+V^{\prime}\right)-\mu\right)}}{12t+6\left(V+V^{\prime}\right)}\, .
\label{eq:ESF}
\end{align}
Energies of the solid phases are obtained as expectation values of
the full Bose--Hubbard Hamiltonian Eq.~\eqref{eq:BH} in states with corresponding fillings $(1/3,2/3,1)$ on the triangular unit cell. Normalizing energies and chemical potential by the scale $\abs{V} + \abs{V^\prime}$, we may write the dimensionless per-site energies of the solid and superfluid (SF) phases as:
\begin{align*}
\tilde{E}_{1/3} & =\frac{-\tilde{\mu}}{3}-\frac{1}{\alpha+1}\\
\tilde{E}_{2/3} & =\frac{-2\tilde{\mu}}{3}+\frac{\text{sgn}(V)\alpha-2}{\alpha+1}\\
\tilde{E}_{1} & =-\tilde{\mu}+3\frac{\text{sgn}(V)\alpha-1}{\alpha+1}\\
\tilde{E}_{SF} & =-\frac{\left(6\tilde{t}+\tilde{\mu}\right)^{2}(\alpha+1)}{24\tilde{t}(\alpha+1)+12\ \qty[\text{sgn}(V)\alpha-1]}
\end{align*}
where we have introduced the notation $\tilde{c} \equiv c/\qty(\abs{V} + \abs{V^\prime})$, and defined $\alpha \equiv \abs{V/V^\prime}$, with $\text{sgn}(\dots)$ the signum function. 

To capture a possible supersolid phase (defined as one having simultaneously broken translational and gauge symmetries), we allow for more degrees of freedom in the mean field decomposition. Considering the triangular unit cell with sites ${\displaystyle A,\,B,\,C}$, we assume
\begin{align*}
\psi_{A}\neq\psi_{B}=\psi_{C}\\
\rho_{A}\neq\rho_{B}=\rho_{C}.
\end{align*}
Decoupling the mean field Hamiltonian in Eq.~\eqref{eq:HMF} for each unit cell yields
\[
H_{MF,\triangle}=H_{t}+H_{V}+H_{V^{\prime}}+H_{\mu},
\]
where 
\begin{align*}
H_{t} &=  -3t\left[2\psi_{B}\left(b_{A}+b_{A}^{\dagger}\right)+\left(\psi_{A}+\psi_{B}\right)\left(b_{B}+b_{B}^{\dagger}\right.\right.\\
&\quad \left.+b_{C}+b_{C}^{\dagger}\right)-4\psi_{A}\psi_{B}-2\psi_{B}^{2}\Bigl]\\
H_{V} &= 3V\left[2\rho_{B}n_{A}+\left(\rho_{A}+\rho_{B}\right)\left(n_{B}+n_{C}\right)-2\rho_{A}\rho_{B}-\rho_{B}^{2}\right]\\
H_{V^{\prime}} &=  3V^{\prime}\left[2\rho_{A}n_{A}+2\rho_{B}\left(n_{B}+n_{C}\right)-\rho_{A}^{2}-2\rho_{B}^{2}\right]\\
H_{\mu} &= -\mu\left(n_{A}+n_{B}+n_{C}\right).
\end{align*}
The energy $E_{SS}$ of the resulting state can be found by numerical solution of 
\[
\partial_{\psi_{A}}E=\partial_{\psi_{B}}E=\partial_{\rho_{A}}E=\partial_{\rho_{B}}E=0
\]
at each point in ${\displaystyle \tilde{\mu}-\tilde{t}}$ phase space. 

The values of $V$ and $V^\prime$ to be used can be determined by Hartree--Fock calculations (see next section) as a function of biaxial strain $\delta$.  Three distinct physical regimes arise in terms of their relative magnitudes as discussed in the main text: 
\[
\begin{array}{cll}
    \vert V\vert\gg\vert V^{\prime}\vert &,\; V>0 & \text{unstrained}\\
    \vert V\vert\simeq\vert V^{\prime}\vert &,\; V>0 & \text{moderate strain}\\
    \vert V\vert\simeq\vert V^{\prime}\vert &,\; V<0\, & \text{large strain}.
\end{array}
\]
Thus, by fixing the ratio $\abs{V/V^\prime}$, mean field phase diagrams can be generated, as shown in Fig.~\ref{fig:MF}.  To obtain a realistic phase diagram for the \he{}-on-graphene system over a range of physical parameters, the full Hartree--Fock results for $t(\delta)$, $V(\delta)$, and $V^\prime(\delta)$ can be used directly, which leads to supplementary Figure~S1.

\subsection{Hartree--Fock}
 
Since multi-particle quantum Monte Carlo simulations are time-consuming, we employed a computationally cheaper method, based on the Hartree--Fock (HF) approximation, 
to compute $V$ and $V'$ in Eq.~\eqref{eq:BH} to be used in the strain-tuned mean field phase diagram.  The HF ansatz for the wavefunction $\Psi$ for $N$ bosons is:
\be
\Psi(\bfr_1,\bfr_2,\ldots,\bfr_N) = \sum_{j(q)} \, \prod_{q=1}^N \phi_{j(q)}(\bfr_q),
\label{eq:HF01}
\ee
where $\bfr_q$ are the 3D coordinates of particle $q$, $j(q)$ is a label of the site where 
particle $q$ is found, and the one-particle quasi-wavefunctions (in what follows we will drop 
``quasi") satisfy the orthonormality conditions
\be
\langle \phi_i | \phi_j\rangle \equiv 
\int d^2\bfr \, \phi^\dagger_i(\bfr)\phi_j(\bfr) = \delta_{i,j}\,,
\label{eq:HW02}
\ee
and the $\dagger$ stands for Hermitian conjugation. 
Employing an approximation $\phi_i(\bfr) \approx \chi(z)\psi_i(\vrg)$, where $\vrg$ is the 2D
coordinate in the plane parallel to the graphene sheet and $z$ is the perpendicular coordinate, 
the 2D-reduced wavefunction can be shown to satisfy HF equations:
\begin{multline}
    -\frac{\hbar^2}{2m}\nabla^2_{\vrg} \psi_i(\vrg) + 
{\mathcal V}_{\graphene}(\vrg) \psi_i(\vrg) \\
+ \sum_{i\neq j} \int \dd{\vrg'}\,\psi_j^\ast(\vrg') \mathcal{V}(\vrg-\vrg') \\
\times \left[ \psi_j(\vrg')\psi_i(\vrg) + \psi_i(\vrg')\psi_j(\vrg) \right] 
= \sum_j E_{ij} \psi_j(\vrg)\,,
 \label{eq:HF03}
\end{multline}
where the Lagrange multipliers $E_{ij}$ are determined by the 2D form of the orthonormality conditions
\eqref{eq:HW02}. Details of these approximations, as well as of the solution method of \eqref{eq:HF03},
were outlined in \cite{Yu:2021tw}. In \eqref{eq:HF03}, ${\mathcal V}_{\rm \graphene}(\vrg)$ is computed as:
 \begin{equation}
     \mathcal{V}_{\graphene}(\vrg) \equiv \expval{\frac{\int \dd{z} 
     \mathcal{V}_{\graphene}(\vrg,z)\rho(\vrg,z)}{\int \dd{z} \rho(\vrg,z)}}
 \label{eq:VHeGrapheneAverage}
 \end{equation}
where $\rho(\vrg,z)$ is the probability density obtained with one-particle (and hence relatively fast) 
QMC simulations and the angle brackets stand for the ensemble average. 
Furthermore, $\mathcal{V}(\vrg)$ is the 2D-reduction (as explained in \cite{Yu:2021tw}) of the interaction
potential $\mathcal{V}(\vrg,z)$ between two helium atoms. In a slight deviation from \cite{Yu:2021tw}, here for the computation of this reduced 2D potential, we used the one-dimensional probability density 
$\rho(z) \equiv \int \dd{\vrg} \rho(\vrg,z)$, with the $\rho(\vrg,z)$ as defined above. 
Then, parameter $V$ in \eqref{eq:BH} is computed as:
\be
V = \int\int \dd{\vrg}\,\dd{\vrg'} \, |\psi_i(\vrg)|^2 \mathcal{V}(\vrg-\vrg') 
|\psi_j(\vrg')|^2,
\label{eq:BHV}
\ee
where $i$ and $j$ are the indices of the two nearest-neighbor graphene cells. Parameter $V'$ is defined
similarly, but for the next-nearest neighbors. Finally, parameter $t$ is computed as described in 
\cite{Yu:2021tw}, using one-particle Wannier functions for a single helium atom over the graphene sheet. 

\subsection{Quantum Monte Carlo}
\label{subsec:QMC}

The strained graphene plus \he{} system described by the Hamiltonian in Eq.~\eqref{eq:Ham} was simulated using a stochastically exact quantum Monte Carlo (QMC) algorithm exploiting path integrals \cite{Ceperley:1995gr,Boninsegni:2006ed,Yu:2021tw}.   Finite temperature expectation values of observables $\mathcal{O}$ were sampled via 
\begin{equation}
    \expval{{\mathcal{O}}}  = \frac{1}{\mathcal{Z}} \Tr(\mathcal{O}\ \mathrm{e}^{-\beta {{H}}})
\label{eq:thermal_expectation_value}
\end{equation}
where $\beta = 1/k_{\rm {B}} T$ is the inverse temperature, $k_{\rm B}$ is the Boltzmann constant, and the partition function $\mathcal{Z} = \Tr\mathrm{e}^{-\beta {{H}}}$ can be written as a sum of discrete imaginary time paths (worldlines) over the set of all permutations $\mathcal{P}$ of the first quantized labels of the $N$ indistinguishable $^4$He atoms. Algorithmic details have been reported elsewhere (e.g.\ Refs.~\cite{Nichols:2020of,Yu:2021tw,DelMaestro:2022wm}) and access to the QMC software is described in the Code Availability Section.

\subsubsection{Simulation Cell}

\begin{figure}
\includegraphics[width=\columnwidth]{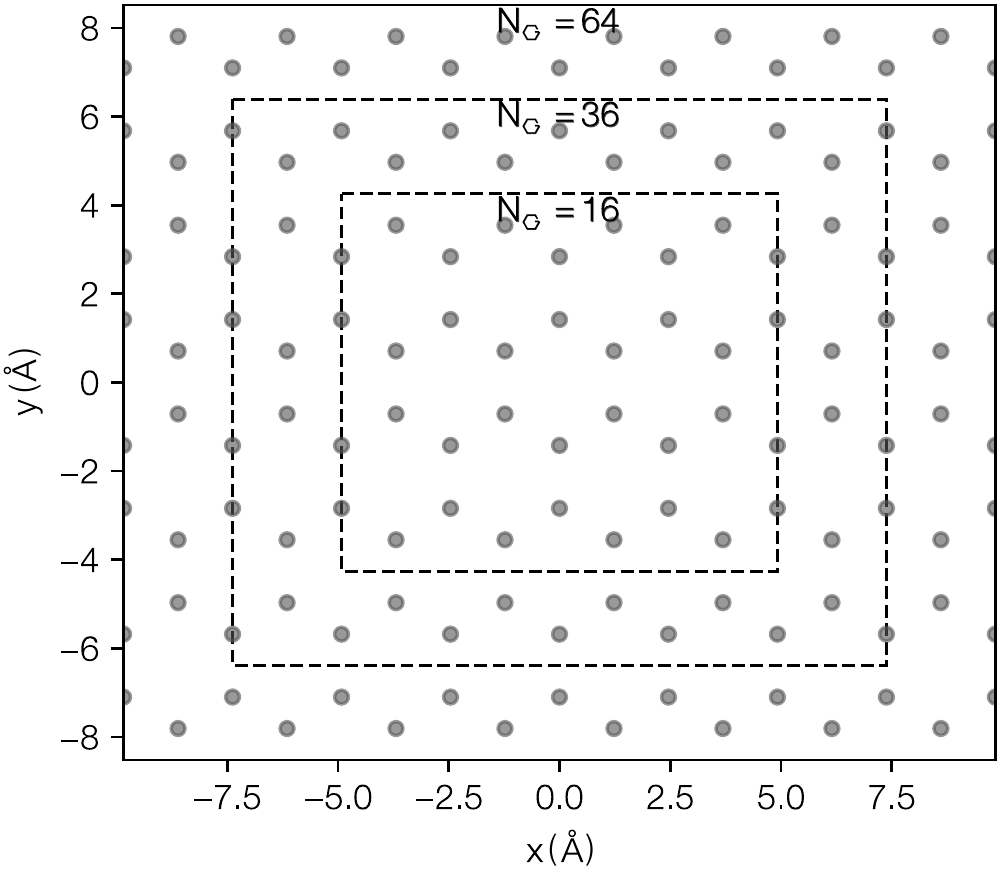}
\caption{\textbf{Finite size simulation cell.} Three system sizes were used for finite size scaling where periodic boundary conditions restrict the box dimensions to be a multiple of lattice vectors in the $x$ and $y$ directions.  The resulting box size is a function of strain $\delta$ and $N_\graphene$ refers to the number of triangular lattice adsorption sites in the simulation cell.}
\label{fig:boxes}
\end{figure}

The simulation cell is defined by a rectangular prism of dimensions $L_x \times L_y \times L_z$ where $L_x$ and $L_y$ are chosen such that the strained graphene sheet is compatible with periodic boundary conditions in the $x$ and $y$ directions.  A membrane with 
$N_{\graphene}=2N_{x}N_{y}$ triangular lattice adsorption sites
requires that $L_x = a_0(1+\delta)\sqrt{3}N_{x}$ for the zigzag direction and $L_y = 3a_0(1+\delta) N_{y}$ for the armchair direction.  Three of the box sizes corresponding to different numbers of adsorption sites used for finite size scaling our QMC results to the thermodynamic limit are shown in Fig.~\ref{fig:boxes}.
The strained graphene membrane is frozen in place at $z=0$ with lattice ($\vb*{a}$) and basis ($\vb*{b}$) vectors:
\begin{equation}
\begin{aligned}
    \vb*{a}_1(\delta) &= \frac{a_0(1+\delta)}{2}\qty(\sqrt{3},3), & \vb*{b}_1(\delta) &= \frac{a_0(1+\delta)}{2}\qty(\sqrt{3},1) \\
    \vb*{a}_2(\delta) &= \frac{a_0(1+\delta)}{2}\qty(-\sqrt{3},3), & \vb*{b}_2(\delta) &= a_0(1+\delta)\qty(0,1) 
\label{eq:grapheneBasis}
\end{aligned}
\end{equation}
where $a_0 \simeq \SI{1.42}{\angstrom}$ is the bare carbon--carbon distance and $0 \le \delta \le 1$ represents its increase under isotropic strain.  Motion in the $z$ direction is restricted via a hard wall placed at $z = L_z = \SI{10}{\angstrom}$, chosen to reproduce bulk multi-layer adsorption phenomena \cite{Yu:2021tw}.

The resulting empirical interaction potential  $\mathcal{V}_{\graphene}$ between \he{} and the strained graphene is computed by assuming a superposition of 6--12 Lennard--Jones potentials between carbon and helium \cite{Steele:1973fo}:
\begin{multline}
    \!\!\!\!\mathcal{V}_{\graphene}(\vb*{r}_i) =  \frac{8\pi\varepsilon(\delta)\sigma^2(\delta)}{3\sqrt{3}a_0^2(1+\delta)^2} \left\{
\qty[\frac{2}{5}\qty(\frac{\sigma(\delta)}{z_i})^{10}\!\!\!\!-\qty(\frac{\sigma(\delta)}{z_i})^{4}] \right. \\ 
    + \sum_{\vb*{g}\ne 0}\sum_{\ell=1}^{2} \mathrm{e}^{\imath \vb*{g}(\delta)\cdot[\vrg_i-\vb*{b}_\ell(\delta)]} \left[\frac{1}{60}\qty(\frac{g(\delta)\sigma(\delta)^2}{2z_i})^{5}K_5(g(\delta)z_i) \right.  \\
\left.\left. - \qty(\frac{g(\delta)\sigma(\delta)^2}{2z_i})^{2}K_2(g(\delta)z_i)\right]  \right\}\, ,
\label{eq:HeGraphene}
\end{multline}
where $\sigma(\delta)$ and $\varepsilon(\delta)$ are strain-dependent Lennard--Jones parameters that have been computed via the method described in Ref.~\cite{Nichols:2016hd} with values and tabulated potentials (up to $\delta = 0.3$) available online \cite{nichols_nathan_s_2021_6574043}.  In Eq.~\eqref{eq:HeGraphene}, $\vrg_i = (x_i,y_i)$ are the coordinates of a $^4$He atom in the $xy$-plane, and  $\vb*{g}(\delta) = n_1 \vb*{G}_1(\delta) + n_2 \vb*{G}_2(\delta)$ are the reciprocal lattice vectors with magnitude $g(\delta) \equiv \abs{\vb*{g}(\delta)}$ where $n_1,n_2 \in \mathds{Z}$,
\begin{align}
    \vb*{G}_1(\delta) &= \frac{2\pi}{3a_0(1+\delta)}\qty(\sqrt{3},1)\, , \nonumber \\
    \vb*{G}_2(\delta) &= \frac{2\pi}{3a_0(1+\delta)}\qty(-\sqrt{3},1)
\label{eq:grapheneG}
\end{align}
and $K_n$ are modified Bessel functions which decay as $\exp(-gz_i)$ at large argument.

\subsubsection{Observables}
To map out the phase diagram reported in Fig.~\ref{fig:pimcPD}, we have computed a number of observables obtained via quantum Monte Carlo estimators.  The total number of particles $N$ can fluctuate in the grand canonical ensemble at fixed temperature $T$ and chemical potential $\mu$ leading to an average value $\expval{N}$ and filling fraction:
\begin{equation}
    n \equiv \frac{\expval{N}}{N_{\graphene}}
\end{equation}
where $N_{\graphene}$ is set by the geometry of the simulation cell, and utilizing the fact that all atoms are adsorbed.   The density of \he{} is given by:
\begin{equation}
    \rho(\vb*{r}) = \expval{\frac{1}{N} \sum_{i=1}^N \delta(\vb*{r}-\vb*{r}_i)}
\end{equation}
where $\delta(\dots)$ is the Dirac delta-function. The planar density of \he{} adsorbed to the graphene can be computed by integrating over $z$: $\rho(x,y) = \int dz \rho(\vb*{r})$ and its resulting compressibility is given by the usual fluctuation measure:
\begin{equation}
    \kappa = \frac{\expval{N^2}-\expval{N}^2}{k_{\rm B} T L_x L_y}\, .
\end{equation}
Finally, the superfluid density $\rho_s$ is related to the response of the free energy to a boundary phase twist \cite{Fisher:1973zm} which can be captured in QMC via the topological winding number $\vb*{W}$ of particle worldlines around the simulation cell \cite{Pollock:1987ta,Prokofev:2000ei,Rousseau:2014pv}:
\begin{equation}
    \rho_s = \frac{m_4^2}{2\hbar^2 \beta L_x L_y} \qty(L_x^2 \expval{W_x^2} + L_y^2 \expval{W_y}^2)
\end{equation}
where
\begin{equation}
    {W}_x = \frac{1}{L_x} \sum_{i=1}^{N} \int_0^{\hbar \beta} \dd{\tau} \qty[\dv{x_i(\tau)}{\tau}]\, .
\end{equation}
with $m_4$ the mass of a \he{} atom and $x_i(\tau)$ is the $x$-coordinate of the imaginary time wordline corresponding to atom $i$.

\subsubsection{Simulation Details and Finite Size Scaling}

Quantum Monte Carlo calculations were performed using open source software \cite{pimczm} for $T = \SIrange{0.5}{2.0}{\kelvin}$ and chemical potentials $\mu / k_{\rm B}$ from $\SIrange{-129}{-41}{\kelvin}$ at four system sizes corresponding to $N_{\graphene} = 16,36,64,144$ triangular lattice adsorption sites to obtain particle configurations at values of the isotropic strain $\delta=0$ (unstrained) to $\delta = 0.3$ (strongly strained). The imaginary time step was fixed at $k_{\rm B}\tau = \SI{0.00313}{K^{-1}}$ such that any systematic effects due to Trotterization are smaller than statistical sampling errors.

By searching for stable plateaus in the filling fraction $n$ at different values of $\delta$ that correspond to vanishing compressibility, we identified commensurate insulating phases corresponding to fillings of $n=1/3, 7/6$, and $1$.   The vacuum phase boundary in Fig.~\ref{fig:pimcPD} corresponds to the line denoting a non-zero expectation value $\expval{N} \ge 0$.

While particle configurations are reported at fixed system sizes, superfluid and particle densities were obtained via a finite size scaling procedure at each temperature to extrapolate to the thermodynamic limit for the cell sizes depicted in Fig.~\ref{fig:boxes}.  Details are included in Supplemental Figure~2, where we have assumed the finite size scaling forms $n(N) = n\vert_\infty + O(1/N)$ and $\rho_s(N) = \rho_s\vert_\infty + O(1/\sqrt{N})$. The superfluid phase boundary in Fig.~\ref{fig:pimcPD} was determined by performing this finite size scaling procedure at 65 $(\delta,\mu)$ points and identifying as superfluid any point where $\rho_s$ persists to the thermodynamic limit for temperatures greater than $T = \SI{0.5}{\kelvin}$ (the base $T$ in our quantum Monte Carlo study).  

Error bars on composite estimators (such as the compressibility) were estimated via jackknife sampling \cite{Young:2012ea}.

\section{Data Availability}
The raw quantum Monte Calro simulation data set is available at \url{https://zenodo.org/record/7271852} \cite{Zenodo:2022} while the processed data can be found online \url{https://github.com/DelMaestroGroup/papers-code-Superfluid4HeStrainGraphene} \cite{sang_wook_kim_2022_7294692}. 

. 

\section{Code Availability}
The code and scripts used to process data and generate all figures in this paper are available online at \url{https://github.com/DelMaestroGroup/papers-code-Superfluid4HeStrainGraphene} \cite{sang_wook_kim_2022_7294692}. The path integral quantum Monte Carlo software used to generate all raw data is available online at \url{https://github.com/DelMaestroGroup/pimc} \cite{pimczm}.

\section{Acknowledgments}

This work was supported by NASA grant number 80NSSC19M0143.  
Computational resources were provided by the NASA High-End Computing (HEC) Program through the NASA Advanced Supercomputing (NAS) Division at Ames Research Center. 


\nocite{apsrev42Control}
\bibliographystyle{apsrev4-2}
\bibliography{refs}

\begin{thebibliography}{69}%
\makeatletter
\providecommand \@ifxundefined [1]{%
 \@ifx{#1\undefined}
}%
\providecommand \@ifnum [1]{%
 \ifnum #1\expandafter \@firstoftwo
 \else \expandafter \@secondoftwo
 \fi
}%
\providecommand \@ifx [1]{%
 \ifx #1\expandafter \@firstoftwo
 \else \expandafter \@secondoftwo
 \fi
}%
\providecommand \natexlab [1]{#1}%
\providecommand \enquote  [1]{``#1''}%
\providecommand \bibnamefont  [1]{#1}%
\providecommand \bibfnamefont [1]{#1}%
\providecommand \citenamefont [1]{#1}%
\providecommand \href@noop [0]{\@secondoftwo}%
\providecommand \href [0]{\begingroup \@sanitize@url \@href}%
\providecommand \@href[1]{\@@startlink{#1}\@@href}%
\providecommand \@@href[1]{\endgroup#1\@@endlink}%
\providecommand \@sanitize@url [0]{\catcode `\\12\catcode `\$12\catcode
  `\&12\catcode `\#12\catcode `\^12\catcode `\_12\catcode `\%12\relax}%
\providecommand \@@startlink[1]{}%
\providecommand \@@endlink[0]{}%
\providecommand \url  [0]{\begingroup\@sanitize@url \@url }%
\providecommand \@url [1]{\endgroup\@href {#1}{\urlprefix }}%
\providecommand \urlprefix  [0]{URL }%
\providecommand \Eprint [0]{\href }%
\providecommand \doibase [0]{https://doi.org/}%
\providecommand \selectlanguage [0]{\@gobble}%
\providecommand \bibinfo  [0]{\@secondoftwo}%
\providecommand \bibfield  [0]{\@secondoftwo}%
\providecommand \translation [1]{[#1]}%
\providecommand \BibitemOpen [0]{}%
\providecommand \bibitemStop [0]{}%
\providecommand \bibitemNoStop [0]{.\EOS\space}%
\providecommand \EOS [0]{\spacefactor3000\relax}%
\providecommand \BibitemShut  [1]{\csname bibitem#1\endcsname}%
\let\auto@bib@innerbib\@empty
\bibitem [{\citenamefont {Altman}\ \emph {et~al.}(2021)\citenamefont {Altman},
  \citenamefont {Brown}, \citenamefont {Carleo}, \citenamefont {Carr},
  \citenamefont {Demler}, \citenamefont {Chin}, \citenamefont {DeMarco},
  \citenamefont {Economou}, \citenamefont {Eriksson}, \citenamefont {Fu},
  \citenamefont {Greiner}, \citenamefont {Hazzard}, \citenamefont {Hulet},
  \citenamefont {Koll{\'{a}}r}, \citenamefont {Lev}, \citenamefont {Lukin},
  \citenamefont {Ma}, \citenamefont {Mi}, \citenamefont {Misra}, \citenamefont
  {Monroe}, \citenamefont {Murch}, \citenamefont {Nazario}, \citenamefont {Ni},
  \citenamefont {Potter}, \citenamefont {Roushan}, \citenamefont {Saffman},
  \citenamefont {Schleier-Smith}, \citenamefont {Siddiqi}, \citenamefont
  {Simmonds}, \citenamefont {Singh}, \citenamefont {Spielman}, \citenamefont
  {Temme}, \citenamefont {Weiss}, \citenamefont {Vu{\v{c}}kovi{\'{c}}},
  \citenamefont {Vuleti{\'{c}}}, \citenamefont {Ye},\ and\ \citenamefont
  {Zwierlein}}]{Altman:2021qm}%
  \BibitemOpen
  \bibfield  {author} {\bibinfo {author} {\bibfnamefont {E.}~\bibnamefont
  {Altman}}, \bibinfo {author} {\bibfnamefont {K.~R.}\ \bibnamefont {Brown}},
  \bibinfo {author} {\bibfnamefont {G.}~\bibnamefont {Carleo}}, \bibinfo
  {author} {\bibfnamefont {L.~D.}\ \bibnamefont {Carr}}, \bibinfo {author}
  {\bibfnamefont {E.}~\bibnamefont {Demler}}, \bibinfo {author} {\bibfnamefont
  {C.}~\bibnamefont {Chin}}, \bibinfo {author} {\bibfnamefont {B.}~\bibnamefont
  {DeMarco}}, \bibinfo {author} {\bibfnamefont {S.~E.}\ \bibnamefont
  {Economou}}, \bibinfo {author} {\bibfnamefont {M.~A.}\ \bibnamefont
  {Eriksson}}, \bibinfo {author} {\bibfnamefont {K.-M.~C.}\ \bibnamefont {Fu}},
  \bibinfo {author} {\bibfnamefont {M.}~\bibnamefont {Greiner}}, \bibinfo
  {author} {\bibfnamefont {K.~R.~A.}\ \bibnamefont {Hazzard}}, \bibinfo
  {author} {\bibfnamefont {R.~G.}\ \bibnamefont {Hulet}}, \bibinfo {author}
  {\bibfnamefont {A.~J.}\ \bibnamefont {Koll{\'{a}}r}}, \bibinfo {author}
  {\bibfnamefont {B.~L.}\ \bibnamefont {Lev}}, \bibinfo {author} {\bibfnamefont
  {M.~D.}\ \bibnamefont {Lukin}}, \bibinfo {author} {\bibfnamefont
  {R.}~\bibnamefont {Ma}}, \bibinfo {author} {\bibfnamefont {X.}~\bibnamefont
  {Mi}}, \bibinfo {author} {\bibfnamefont {S.}~\bibnamefont {Misra}}, \bibinfo
  {author} {\bibfnamefont {C.}~\bibnamefont {Monroe}}, \bibinfo {author}
  {\bibfnamefont {K.}~\bibnamefont {Murch}}, \bibinfo {author} {\bibfnamefont
  {Z.}~\bibnamefont {Nazario}}, \bibinfo {author} {\bibfnamefont {K.-K.}\
  \bibnamefont {Ni}}, \bibinfo {author} {\bibfnamefont {A.~C.}\ \bibnamefont
  {Potter}}, \bibinfo {author} {\bibfnamefont {P.}~\bibnamefont {Roushan}},
  \bibinfo {author} {\bibfnamefont {M.}~\bibnamefont {Saffman}}, \bibinfo
  {author} {\bibfnamefont {M.}~\bibnamefont {Schleier-Smith}}, \bibinfo
  {author} {\bibfnamefont {I.}~\bibnamefont {Siddiqi}}, \bibinfo {author}
  {\bibfnamefont {R.}~\bibnamefont {Simmonds}}, \bibinfo {author}
  {\bibfnamefont {M.}~\bibnamefont {Singh}}, \bibinfo {author} {\bibfnamefont
  {I.~B.}\ \bibnamefont {Spielman}}, \bibinfo {author} {\bibfnamefont
  {K.}~\bibnamefont {Temme}}, \bibinfo {author} {\bibfnamefont {D.~S.}\
  \bibnamefont {Weiss}}, \bibinfo {author} {\bibfnamefont {J.}~\bibnamefont
  {Vu{\v{c}}kovi{\'{c}}}}, \bibinfo {author} {\bibfnamefont {V.}~\bibnamefont
  {Vuleti{\'{c}}}}, \bibinfo {author} {\bibfnamefont {J.}~\bibnamefont {Ye}},\
  and\ \bibinfo {author} {\bibfnamefont {M.}~\bibnamefont {Zwierlein}},\
  }\bibfield  {title} {\bibinfo {title} {{Q}uantum {S}imulators:
  {A}rchitectures and {O}pportunities},\ }\href
  {https://doi.org/10.1103/prxquantum.2.017003} {\bibfield  {journal} {\bibinfo
   {journal} {{PRX} Quantum}\ }\textbf {\bibinfo {volume} {2}},\ \bibinfo
  {pages} {017003} (\bibinfo {year} {2021})}\BibitemShut {NoStop}%
\bibitem [{\citenamefont {Bloch}\ \emph {et~al.}(2008)\citenamefont {Bloch},
  \citenamefont {Dalibard},\ and\ \citenamefont {Zwerger}}]{Bloch:2008}%
  \BibitemOpen
  \bibfield  {author} {\bibinfo {author} {\bibfnamefont {I.}~\bibnamefont
  {Bloch}}, \bibinfo {author} {\bibfnamefont {J.}~\bibnamefont {Dalibard}},\
  and\ \bibinfo {author} {\bibfnamefont {W.}~\bibnamefont {Zwerger}},\
  }\bibfield  {title} {\bibinfo {title} {Many-body physics with ultracold
  gases},\ }\href {https://doi.org/10.1103/RevModPhys.80.885} {\bibfield
  {journal} {\bibinfo  {journal} {Rev. Mod. Phys.}\ }\textbf {\bibinfo {volume}
  {80}},\ \bibinfo {pages} {885} (\bibinfo {year} {2008})}\BibitemShut
  {NoStop}%
\bibitem [{\citenamefont {Lewenstein}\ \emph {et~al.}(2007)\citenamefont
  {Lewenstein}, \citenamefont {Sanpera}, \citenamefont {Ahufinger},
  \citenamefont {Damski}, \citenamefont {Sen(De)},\ and\ \citenamefont
  {Sen}}]{Lewenstein}%
  \BibitemOpen
  \bibfield  {author} {\bibinfo {author} {\bibfnamefont {M.}~\bibnamefont
  {Lewenstein}}, \bibinfo {author} {\bibfnamefont {A.}~\bibnamefont {Sanpera}},
  \bibinfo {author} {\bibfnamefont {V.}~\bibnamefont {Ahufinger}}, \bibinfo
  {author} {\bibfnamefont {B.}~\bibnamefont {Damski}}, \bibinfo {author}
  {\bibfnamefont {A.}~\bibnamefont {Sen(De)}},\ and\ \bibinfo {author}
  {\bibfnamefont {U.}~\bibnamefont {Sen}},\ }\bibfield  {title} {\bibinfo
  {title} {Ultracold atomic gases in optical lattices: mimicking condensed
  matter physics and beyond},\ }\href
  {https://doi.org/10.1080/00018730701223200} {\bibfield  {journal} {\bibinfo
  {journal} {Adv. Phys.}\ }\textbf {\bibinfo {volume} {56}},\ \bibinfo {pages}
  {243} (\bibinfo {year} {2007})}\BibitemShut {NoStop}%
\bibitem [{\citenamefont {Zhang}\ \emph {et~al.}(2018)\citenamefont {Zhang},
  \citenamefont {Zhu}, \citenamefont {Zhao}, \citenamefont {Yan},\ and\
  \citenamefont {Zhu}}]{Zhang2018}%
  \BibitemOpen
  \bibfield  {author} {\bibinfo {author} {\bibfnamefont {D.-W.}\ \bibnamefont
  {Zhang}}, \bibinfo {author} {\bibfnamefont {Y.-Q.}\ \bibnamefont {Zhu}},
  \bibinfo {author} {\bibfnamefont {Y.~X.}\ \bibnamefont {Zhao}}, \bibinfo
  {author} {\bibfnamefont {H.}~\bibnamefont {Yan}},\ and\ \bibinfo {author}
  {\bibfnamefont {S.-L.}\ \bibnamefont {Zhu}},\ }\bibfield  {title} {\bibinfo
  {title} {Topological quantum matter with cold atoms},\ }\href
  {https://doi.org/10.1080/00018732.2019.1594094} {\bibfield  {journal}
  {\bibinfo  {journal} {Adv. Phys.}\ }\textbf {\bibinfo {volume} {67}},\
  \bibinfo {pages} {253} (\bibinfo {year} {2018})}\BibitemShut {NoStop}%
\bibitem [{\citenamefont {Castro~Neto}\ \emph {et~al.}(2009)\citenamefont
  {Castro~Neto}, \citenamefont {Guinea}, \citenamefont {Peres}, \citenamefont
  {Novoselov},\ and\ \citenamefont {Geim}}]{Antonio}%
  \BibitemOpen
  \bibfield  {author} {\bibinfo {author} {\bibfnamefont {A.~H.}\ \bibnamefont
  {Castro~Neto}}, \bibinfo {author} {\bibfnamefont {F.}~\bibnamefont {Guinea}},
  \bibinfo {author} {\bibfnamefont {N.~M.~R.}\ \bibnamefont {Peres}}, \bibinfo
  {author} {\bibfnamefont {K.~S.}\ \bibnamefont {Novoselov}},\ and\ \bibinfo
  {author} {\bibfnamefont {A.~K.}\ \bibnamefont {Geim}},\ }\bibfield  {title}
  {\bibinfo {title} {The electronic properties of graphene},\ }\href
  {https://doi.org/10.1103/RevModPhys.81.109} {\bibfield  {journal} {\bibinfo
  {journal} {Rev. Mod. Phys.}\ }\textbf {\bibinfo {volume} {81}},\ \bibinfo
  {pages} {109} (\bibinfo {year} {2009})}\BibitemShut {NoStop}%
\bibitem [{\citenamefont {Geim}\ and\ \citenamefont
  {Grigorieva}(2013)}]{vdwgeim}%
  \BibitemOpen
  \bibfield  {author} {\bibinfo {author} {\bibfnamefont {A.~K.}\ \bibnamefont
  {Geim}}\ and\ \bibinfo {author} {\bibfnamefont {I.}~\bibnamefont
  {Grigorieva}},\ }\bibfield  {title} {\bibinfo {title} {{Van der Waals
  heterostructures}},\ }\href {http://dx.doi.org/10.1038/nature12385}
  {\bibfield  {journal} {\bibinfo  {journal} {Nature}\ }\textbf {\bibinfo
  {volume} {499}},\ \bibinfo {pages} {419} (\bibinfo {year}
  {2013})}\BibitemShut {NoStop}%
\bibitem [{\citenamefont {Semeghini}\ \emph {et~al.}(2021)\citenamefont
  {Semeghini}, \citenamefont {Levine}, \citenamefont {Keesling}, \citenamefont
  {Ebadi}, \citenamefont {Wang}, \citenamefont {Bluvstein}, \citenamefont
  {Verresen}, \citenamefont {Pichler}, \citenamefont {Kalinowski},
  \citenamefont {Samajdar}, \citenamefont {Omran}, \citenamefont {Sachdev},
  \citenamefont {Vishwanath}, \citenamefont {Greiner}, \citenamefont
  {Vuleti{\'{c}}},\ and\ \citenamefont {Lukin}}]{Semeghini:2021dl}%
  \BibitemOpen
  \bibfield  {author} {\bibinfo {author} {\bibfnamefont {G.}~\bibnamefont
  {Semeghini}}, \bibinfo {author} {\bibfnamefont {H.}~\bibnamefont {Levine}},
  \bibinfo {author} {\bibfnamefont {A.}~\bibnamefont {Keesling}}, \bibinfo
  {author} {\bibfnamefont {S.}~\bibnamefont {Ebadi}}, \bibinfo {author}
  {\bibfnamefont {T.~T.}\ \bibnamefont {Wang}}, \bibinfo {author}
  {\bibfnamefont {D.}~\bibnamefont {Bluvstein}}, \bibinfo {author}
  {\bibfnamefont {R.}~\bibnamefont {Verresen}}, \bibinfo {author}
  {\bibfnamefont {H.}~\bibnamefont {Pichler}}, \bibinfo {author} {\bibfnamefont
  {M.}~\bibnamefont {Kalinowski}}, \bibinfo {author} {\bibfnamefont
  {R.}~\bibnamefont {Samajdar}}, \bibinfo {author} {\bibfnamefont
  {A.}~\bibnamefont {Omran}}, \bibinfo {author} {\bibfnamefont
  {S.}~\bibnamefont {Sachdev}}, \bibinfo {author} {\bibfnamefont
  {A.}~\bibnamefont {Vishwanath}}, \bibinfo {author} {\bibfnamefont
  {M.}~\bibnamefont {Greiner}}, \bibinfo {author} {\bibfnamefont
  {V.}~\bibnamefont {Vuleti{\'{c}}}},\ and\ \bibinfo {author} {\bibfnamefont
  {M.~D.}\ \bibnamefont {Lukin}},\ }\bibfield  {title} {\bibinfo {title}
  {{P}robing topological spin liquids on a programmable quantum simulator},\
  }\href {https://doi.org/10.1126/science.abi8794} {\bibfield  {journal}
  {\bibinfo  {journal} {Science}\ }\textbf {\bibinfo {volume} {374}},\ \bibinfo
  {pages} {1242} (\bibinfo {year} {2021})}\BibitemShut {NoStop}%
\bibitem [{\citenamefont {Ma}\ \emph {et~al.}(2019)\citenamefont {Ma},
  \citenamefont {Saxberg}, \citenamefont {Owens}, \citenamefont {Leung},
  \citenamefont {Lu}, \citenamefont {Simon},\ and\ \citenamefont
  {Schuster}}]{Ma:2019zp}%
  \BibitemOpen
  \bibfield  {author} {\bibinfo {author} {\bibfnamefont {R.}~\bibnamefont
  {Ma}}, \bibinfo {author} {\bibfnamefont {B.}~\bibnamefont {Saxberg}},
  \bibinfo {author} {\bibfnamefont {C.}~\bibnamefont {Owens}}, \bibinfo
  {author} {\bibfnamefont {N.}~\bibnamefont {Leung}}, \bibinfo {author}
  {\bibfnamefont {Y.}~\bibnamefont {Lu}}, \bibinfo {author} {\bibfnamefont
  {J.}~\bibnamefont {Simon}},\ and\ \bibinfo {author} {\bibfnamefont {D.~I.}\
  \bibnamefont {Schuster}},\ }\bibfield  {title} {\bibinfo {title} {{A}
  dissipatively stabilized {M}ott insulator of photons},\ }\href
  {https://doi.org/10.1038/s41586-019-0897-9} {\bibfield  {journal} {\bibinfo
  {journal} {Nature}\ }\textbf {\bibinfo {volume} {566}},\ \bibinfo {pages}
  {51} (\bibinfo {year} {2019})}\BibitemShut {NoStop}%
\bibitem [{\citenamefont {Ming}\ \emph {et~al.}(2022)\citenamefont {Ming},
  \citenamefont {Wu}, \citenamefont {Chen}, \citenamefont {Wang}, \citenamefont
  {Mai}, \citenamefont {Maier}, \citenamefont {Strockoz}, \citenamefont
  {Venderbos}, \citenamefont {Gonzalez}, \citenamefont {Ortega}, \citenamefont
  {Johnston},\ and\ \citenamefont {Weitering}}]{Ming:2022zm}%
  \BibitemOpen
  \bibfield  {author} {\bibinfo {author} {\bibfnamefont {F.}~\bibnamefont
  {Ming}}, \bibinfo {author} {\bibfnamefont {X.}~\bibnamefont {Wu}}, \bibinfo
  {author} {\bibfnamefont {C.}~\bibnamefont {Chen}}, \bibinfo {author}
  {\bibfnamefont {K.~D.}\ \bibnamefont {Wang}}, \bibinfo {author}
  {\bibfnamefont {P.}~\bibnamefont {Mai}}, \bibinfo {author} {\bibfnamefont
  {T.~A.}\ \bibnamefont {Maier}}, \bibinfo {author} {\bibfnamefont
  {J.}~\bibnamefont {Strockoz}}, \bibinfo {author} {\bibfnamefont {J.~W.~F.}\
  \bibnamefont {Venderbos}}, \bibinfo {author} {\bibfnamefont {C.}~\bibnamefont
  {Gonzalez}}, \bibinfo {author} {\bibfnamefont {J.}~\bibnamefont {Ortega}},
  \bibinfo {author} {\bibfnamefont {S.}~\bibnamefont {Johnston}},\ and\
  \bibinfo {author} {\bibfnamefont {H.~H.}\ \bibnamefont {Weitering}},\ }\href
  {https://doi.org/10.48550/arxiv.2210.06273} {\bibinfo {title} {{E}vidence for
  chiral superconductivity on a silicon surface}} (\bibinfo {year} {2022}),\
  \Eprint {https://arxiv.org/abs/arXiv:2210.06273} {arXiv:2210.06273}
  \BibitemShut {NoStop}%
\bibitem [{\citenamefont {Kreisel}\ \emph {et~al.}(2021)\citenamefont
  {Kreisel}, \citenamefont {Hyart},\ and\ \citenamefont
  {Rosenow}}]{Kreisel:2021xx}%
  \BibitemOpen
  \bibfield  {author} {\bibinfo {author} {\bibfnamefont {A.}~\bibnamefont
  {Kreisel}}, \bibinfo {author} {\bibfnamefont {T.}~\bibnamefont {Hyart}},\
  and\ \bibinfo {author} {\bibfnamefont {B.}~\bibnamefont {Rosenow}},\
  }\bibfield  {title} {\bibinfo {title} {Tunable topological states hosted by
  unconventional superconductors with adatoms},\ }\href
  {https://doi.org/10.1103/PhysRevResearch.3.033049} {\bibfield  {journal}
  {\bibinfo  {journal} {Phys. Rev. Research}\ }\textbf {\bibinfo {volume}
  {3}},\ \bibinfo {pages} {033049} (\bibinfo {year} {2021})}\BibitemShut
  {NoStop}%
\bibitem [{\citenamefont {{Del Maestro}}\ \emph {et~al.}(2021)\citenamefont
  {{Del Maestro}}, \citenamefont {Wexler}, \citenamefont {Vanegas},
  \citenamefont {Lakoba},\ and\ \citenamefont {Kotov}}]{DelMaestro:2021kc}%
  \BibitemOpen
  \bibfield  {author} {\bibinfo {author} {\bibfnamefont {A.}~\bibnamefont {{Del
  Maestro}}}, \bibinfo {author} {\bibfnamefont {C.}~\bibnamefont {Wexler}},
  \bibinfo {author} {\bibfnamefont {J.~M.}\ \bibnamefont {Vanegas}}, \bibinfo
  {author} {\bibfnamefont {T.}~\bibnamefont {Lakoba}},\ and\ \bibinfo {author}
  {\bibfnamefont {V.~N.}\ \bibnamefont {Kotov}},\ }\bibfield  {title} {\bibinfo
  {title} {{A} perspective on {C}ollective {P}roperties of {A}toms on 2{D}
  materials},\ }\href {https://doi.org/10.1002/aelm.202100607} {\bibfield
  {journal} {\bibinfo  {journal} {Adv. Electron. Mater.}\ }\textbf {\bibinfo
  {volume} {8}},\ \bibinfo {pages} {2100607} (\bibinfo {year}
  {2021})}\BibitemShut {NoStop}%
\bibitem [{\citenamefont {Yu}\ \emph {et~al.}(2021)\citenamefont {Yu},
  \citenamefont {Lauricella}, \citenamefont {Elsayed}, \citenamefont
  {Shepherd}, \citenamefont {Nichols}, \citenamefont {Lombardi}, \citenamefont
  {Kim}, \citenamefont {Wexler}, \citenamefont {Vanegas}, \citenamefont
  {Lakoba}, \citenamefont {Kotov},\ and\ \citenamefont
  {Del~Maestro}}]{Yu:2021tw}%
  \BibitemOpen
  \bibfield  {author} {\bibinfo {author} {\bibfnamefont {J.}~\bibnamefont
  {Yu}}, \bibinfo {author} {\bibfnamefont {E.}~\bibnamefont {Lauricella}},
  \bibinfo {author} {\bibfnamefont {M.}~\bibnamefont {Elsayed}}, \bibinfo
  {author} {\bibfnamefont {K.}~\bibnamefont {Shepherd}}, \bibinfo {author}
  {\bibfnamefont {N.~S.}\ \bibnamefont {Nichols}}, \bibinfo {author}
  {\bibfnamefont {T.}~\bibnamefont {Lombardi}}, \bibinfo {author}
  {\bibfnamefont {S.~W.}\ \bibnamefont {Kim}}, \bibinfo {author} {\bibfnamefont
  {C.}~\bibnamefont {Wexler}}, \bibinfo {author} {\bibfnamefont {J.~M.}\
  \bibnamefont {Vanegas}}, \bibinfo {author} {\bibfnamefont {T.}~\bibnamefont
  {Lakoba}}, \bibinfo {author} {\bibfnamefont {V.~N.}\ \bibnamefont {Kotov}},\
  and\ \bibinfo {author} {\bibfnamefont {A.}~\bibnamefont {Del~Maestro}},\
  }\bibfield  {title} {\bibinfo {title} {{Two-dimensional Bose-Hubbard model
  for helium on graphene}},\ }\href
  {https://doi.org/10.1103/PhysRevB.103.235414} {\bibfield  {journal} {\bibinfo
   {journal} {Phys. Rev. B}\ }\textbf {\bibinfo {volume} {103}},\ \bibinfo
  {pages} {235414} (\bibinfo {year} {2021})}\BibitemShut {NoStop}%
\bibitem [{\citenamefont {Henkel}\ \emph {et~al.}(1969)\citenamefont {Henkel},
  \citenamefont {Smith},\ and\ \citenamefont {Reppy}}]{Henkel:1969gx}%
  \BibitemOpen
  \bibfield  {author} {\bibinfo {author} {\bibfnamefont {R.}~\bibnamefont
  {Henkel}}, \bibinfo {author} {\bibfnamefont {E.}~\bibnamefont {Smith}},\ and\
  \bibinfo {author} {\bibfnamefont {J.}~\bibnamefont {Reppy}},\ }\bibfield
  {title} {\bibinfo {title} {{Temperature Dependence of the Superfluid Healing
  Length}},\ }\href {https://doi.org/10.1103/PhysRevLett.23.1276} {\bibfield
  {journal} {\bibinfo  {journal} {Phys. Rev. Lett.}\ }\textbf {\bibinfo
  {volume} {23}},\ \bibinfo {pages} {1276} (\bibinfo {year}
  {1969})}\BibitemShut {NoStop}%
\bibitem [{\citenamefont {Agnolet}\ \emph {et~al.}(1989)\citenamefont
  {Agnolet}, \citenamefont {McQueeney},\ and\ \citenamefont
  {Reppy}}]{Agnolet:1989ou}%
  \BibitemOpen
  \bibfield  {author} {\bibinfo {author} {\bibfnamefont {G.}~\bibnamefont
  {Agnolet}}, \bibinfo {author} {\bibfnamefont {D.~F.}\ \bibnamefont
  {McQueeney}},\ and\ \bibinfo {author} {\bibfnamefont {J.~D.}\ \bibnamefont
  {Reppy}},\ }\bibfield  {title} {\bibinfo {title} {{K}osterlitz-{T}houless
  transition in helium films},\ }\href
  {https://doi.org/10.1103/physrevb.39.8934} {\bibfield  {journal} {\bibinfo
  {journal} {Phys. Rev. B}\ }\textbf {\bibinfo {volume} {39}},\ \bibinfo
  {pages} {8934} (\bibinfo {year} {1989})}\BibitemShut {NoStop}%
\bibitem [{\citenamefont {Bretz}\ and\ \citenamefont
  {Dash}(1971)}]{Bretz:1971jo}%
  \BibitemOpen
  \bibfield  {author} {\bibinfo {author} {\bibfnamefont {M.}~\bibnamefont
  {Bretz}}\ and\ \bibinfo {author} {\bibfnamefont {J.}~\bibnamefont {Dash}},\
  }\bibfield  {title} {\bibinfo {title} {{Quasiclassical and Quantum Degenerate
  Helium Monolayers}},\ }\href {https://doi.org/10.1103/physrevlett.26.963}
  {\bibfield  {journal} {\bibinfo  {journal} {Phys. Rev. Lett.}\ }\textbf
  {\bibinfo {volume} {26}},\ \bibinfo {pages} {963 } (\bibinfo {year}
  {1971})}\BibitemShut {NoStop}%
\bibitem [{\citenamefont {Bretz}\ \emph {et~al.}(1973)\citenamefont {Bretz},
  \citenamefont {Dash}, \citenamefont {Hickernell}, \citenamefont {McLean},\
  and\ \citenamefont {Vilches}}]{Bretz:1973ky}%
  \BibitemOpen
  \bibfield  {author} {\bibinfo {author} {\bibfnamefont {M.}~\bibnamefont
  {Bretz}}, \bibinfo {author} {\bibfnamefont {J.~G.}\ \bibnamefont {Dash}},
  \bibinfo {author} {\bibfnamefont {D.~C.}\ \bibnamefont {Hickernell}},
  \bibinfo {author} {\bibfnamefont {E.~O.}\ \bibnamefont {McLean}},\ and\
  \bibinfo {author} {\bibfnamefont {O.~E.}\ \bibnamefont {Vilches}},\
  }\bibfield  {title} {\bibinfo {title} {{P}hases of {H}e$^3$ and {H}e$^4$
  {M}onolayer {F}ilms {A}dsorbed on {B}asal-{P}lane {O}riented {G}raphite},\
  }\href {https://doi.org/10.1103/physreva.8.1589} {\bibfield  {journal}
  {\bibinfo  {journal} {Phys. Rev. A}\ }\textbf {\bibinfo {volume} {8}},\
  \bibinfo {pages} {1589} (\bibinfo {year} {1973})}\BibitemShut {NoStop}%
\bibitem [{\citenamefont {Zimmerli}\ and\ \citenamefont
  {Chan}(1988)}]{Zimmerli:1988ii}%
  \BibitemOpen
  \bibfield  {author} {\bibinfo {author} {\bibfnamefont {G.}~\bibnamefont
  {Zimmerli}}\ and\ \bibinfo {author} {\bibfnamefont {M.~H.~W.}\ \bibnamefont
  {Chan}},\ }\bibfield  {title} {\bibinfo {title} {{Complete wetting of helium
  on graphite}},\ }\href {https://doi.org/10.1103/physrevb.38.8760} {\bibfield
  {journal} {\bibinfo  {journal} {Phys. Rev. B}\ }\textbf {\bibinfo {volume}
  {38}},\ \bibinfo {pages} {8760 } (\bibinfo {year} {1988})}\BibitemShut
  {NoStop}%
\bibitem [{\citenamefont {Greywall}\ and\ \citenamefont
  {Busch}(1991)}]{Greywall:1991ns}%
  \BibitemOpen
  \bibfield  {author} {\bibinfo {author} {\bibfnamefont {D.~S.}\ \bibnamefont
  {Greywall}}\ and\ \bibinfo {author} {\bibfnamefont {P.~A.}\ \bibnamefont
  {Busch}},\ }\bibfield  {title} {\bibinfo {title} {{H}eat capacity of fluid
  monolayers of $^4${H}e},\ }\href
  {https://doi.org/10.1103/physrevlett.67.3535} {\bibfield  {journal} {\bibinfo
   {journal} {Phys. Rev. Lett.}\ }\textbf {\bibinfo {volume} {67}},\ \bibinfo
  {pages} {3535} (\bibinfo {year} {1991})}\BibitemShut {NoStop}%
\bibitem [{\citenamefont {Zimmerli}\ \emph {et~al.}(1992)\citenamefont
  {Zimmerli}, \citenamefont {Mistura},\ and\ \citenamefont
  {Chan}}]{Zimmerli:1992hz}%
  \BibitemOpen
  \bibfield  {author} {\bibinfo {author} {\bibfnamefont {G.}~\bibnamefont
  {Zimmerli}}, \bibinfo {author} {\bibfnamefont {G.}~\bibnamefont {Mistura}},\
  and\ \bibinfo {author} {\bibfnamefont {M.~H.~W.}\ \bibnamefont {Chan}},\
  }\bibfield  {title} {\bibinfo {title} {{Third-sound study of a layered
  superfluid film}},\ }\href {https://doi.org/10.1103/physrevlett.68.60}
  {\bibfield  {journal} {\bibinfo  {journal} {Phys. Rev. Lett.}\ }\textbf
  {\bibinfo {volume} {68}},\ \bibinfo {pages} {60 } (\bibinfo {year}
  {1992})}\BibitemShut {NoStop}%
\bibitem [{\citenamefont {Crowell}\ and\ \citenamefont
  {Reppy}(1996)}]{Crowell:1996kn}%
  \BibitemOpen
  \bibfield  {author} {\bibinfo {author} {\bibfnamefont {P.~A.}\ \bibnamefont
  {Crowell}}\ and\ \bibinfo {author} {\bibfnamefont {J.~D.}\ \bibnamefont
  {Reppy}},\ }\bibfield  {title} {\bibinfo {title} {{S}uperfluidity and film
  structure {in {H}e}$^4$ adsorbed on graphite},\ }\href
  {https://doi.org/10.1103/physrevb.53.2701} {\bibfield  {journal} {\bibinfo
  {journal} {Phys. Rev. B}\ }\textbf {\bibinfo {volume} {53}},\ \bibinfo
  {pages} {2701} (\bibinfo {year} {1996})}\BibitemShut {NoStop}%
\bibitem [{\citenamefont {Ny{\'{e}}ki}\ \emph {et~al.}(1998)\citenamefont
  {Ny{\'{e}}ki}, \citenamefont {Ray}, \citenamefont {Cowan},\ and\
  \citenamefont {Saunders}}]{Nyeki:1998fk}%
  \BibitemOpen
  \bibfield  {author} {\bibinfo {author} {\bibfnamefont {J.}~\bibnamefont
  {Ny{\'{e}}ki}}, \bibinfo {author} {\bibfnamefont {R.}~\bibnamefont {Ray}},
  \bibinfo {author} {\bibfnamefont {B.}~\bibnamefont {Cowan}},\ and\ \bibinfo
  {author} {\bibfnamefont {J.}~\bibnamefont {Saunders}},\ }\bibfield  {title}
  {\bibinfo {title} {{S}uperfluidity of {A}tomically {L}ayered $^4$he
  {F}ilms},\ }\href {https://doi.org/10.1103/physrevlett.81.152} {\bibfield
  {journal} {\bibinfo  {journal} {Phys. Rev. Lett.}\ }\textbf {\bibinfo
  {volume} {81}},\ \bibinfo {pages} {152} (\bibinfo {year} {1998})}\BibitemShut
  {NoStop}%
\bibitem [{\citenamefont {Whitlock}\ \emph {et~al.}(1998)\citenamefont
  {Whitlock}, \citenamefont {Chester},\ and\ \citenamefont
  {Krishnamachari}}]{Whitlock:1998gb}%
  \BibitemOpen
  \bibfield  {author} {\bibinfo {author} {\bibfnamefont {P.~A.}\ \bibnamefont
  {Whitlock}}, \bibinfo {author} {\bibfnamefont {G.~V.}\ \bibnamefont
  {Chester}},\ and\ \bibinfo {author} {\bibfnamefont {B.}~\bibnamefont
  {Krishnamachari}},\ }\bibfield  {title} {\bibinfo {title} {{Monte Carlo
  simulation of a helium film on graphite}},\ }\href
  {https://doi.org/10.1103/physrevb.58.8704} {\bibfield  {journal} {\bibinfo
  {journal} {Phys. Rev. B}\ }\textbf {\bibinfo {volume} {58}},\ \bibinfo
  {pages} {8704 } (\bibinfo {year} {1998})}\BibitemShut {NoStop}%
\bibitem [{\citenamefont {Corboz}\ \emph {et~al.}(2008)\citenamefont {Corboz},
  \citenamefont {Boninsegni}, \citenamefont {Pollet},\ and\ \citenamefont
  {Troyer}}]{Corboz2008cb}%
  \BibitemOpen
  \bibfield  {author} {\bibinfo {author} {\bibfnamefont {P.}~\bibnamefont
  {Corboz}}, \bibinfo {author} {\bibfnamefont {M.}~\bibnamefont {Boninsegni}},
  \bibinfo {author} {\bibfnamefont {L.}~\bibnamefont {Pollet}},\ and\ \bibinfo
  {author} {\bibfnamefont {M.}~\bibnamefont {Troyer}},\ }\bibfield  {title}
  {\bibinfo {title} {Phase diagram of $^{4}${H}e adsorbed on graphite},\ }\href
  {https://doi.org/10.1103/physrevb.78.245414} {\bibfield  {journal} {\bibinfo
  {journal} {Phys. Rev. B}\ }\textbf {\bibinfo {volume} {78}},\ \bibinfo
  {pages} {245414} (\bibinfo {year} {2008})}\BibitemShut {NoStop}%
\bibitem [{\citenamefont {Pierce}\ and\ \citenamefont
  {Manousakis}(2000)}]{Pierce:2000cj}%
  \BibitemOpen
  \bibfield  {author} {\bibinfo {author} {\bibfnamefont {M.~E.}\ \bibnamefont
  {Pierce}}\ and\ \bibinfo {author} {\bibfnamefont {E.}~\bibnamefont
  {Manousakis}},\ }\bibfield  {title} {\bibinfo {title} {{Role of substrate
  corrugation in helium monolayer solidification}},\ }\href
  {https://doi.org/10.1103/physrevb.62.5228} {\bibfield  {journal} {\bibinfo
  {journal} {Phys. Rev. B}\ }\textbf {\bibinfo {volume} {62}},\ \bibinfo
  {pages} {5228 } (\bibinfo {year} {2000})}\BibitemShut {NoStop}%
\bibitem [{\citenamefont {Ahn}\ \emph {et~al.}(2016)\citenamefont {Ahn},
  \citenamefont {Lee},\ and\ \citenamefont {Kwon}}]{Ahn:2016dm}%
  \BibitemOpen
  \bibfield  {author} {\bibinfo {author} {\bibfnamefont {J.}~\bibnamefont
  {Ahn}}, \bibinfo {author} {\bibfnamefont {H.}~\bibnamefont {Lee}},\ and\
  \bibinfo {author} {\bibfnamefont {Y.}~\bibnamefont {Kwon}},\ }\bibfield
  {title} {\bibinfo {title} {{Prediction of stable C7/12 and metastable C4/7
  commensurate solid phases for $^4${H}e on graphite}},\ }\href
  {https://doi.org/10.1103/physrevb.93.064511} {\bibfield  {journal} {\bibinfo
  {journal} {Phys. Rev. B}\ }\textbf {\bibinfo {volume} {93}},\ \bibinfo
  {pages} {064511} (\bibinfo {year} {2016})}\BibitemShut {NoStop}%
\bibitem [{\citenamefont {Wessel}\ and\ \citenamefont
  {Troyer}(2005)}]{Wessel:2005ik}%
  \BibitemOpen
  \bibfield  {author} {\bibinfo {author} {\bibfnamefont {S.}~\bibnamefont
  {Wessel}}\ and\ \bibinfo {author} {\bibfnamefont {M.}~\bibnamefont
  {Troyer}},\ }\bibfield  {title} {\bibinfo {title} {{S}upersolid {H}ard-{C}ore
  {B}osons on the {T}riangular {L}attice},\ }\href
  {https://doi.org/10.1103/physrevlett.95.127205} {\bibfield  {journal}
  {\bibinfo  {journal} {Phys. Rev. Lett.}\ }\textbf {\bibinfo {volume} {95}},\
  \bibinfo {pages} {127205} (\bibinfo {year} {2005})}\BibitemShut {NoStop}%
\bibitem [{\citenamefont {Nakamura}\ \emph {et~al.}(2016)\citenamefont
  {Nakamura}, \citenamefont {Matsui}, \citenamefont {Matsui},\ and\
  \citenamefont {Fukuyama}}]{Nakamura:2016wf}%
  \BibitemOpen
  \bibfield  {author} {\bibinfo {author} {\bibfnamefont {S.}~\bibnamefont
  {Nakamura}}, \bibinfo {author} {\bibfnamefont {K.}~\bibnamefont {Matsui}},
  \bibinfo {author} {\bibfnamefont {T.}~\bibnamefont {Matsui}},\ and\ \bibinfo
  {author} {\bibfnamefont {H.}~\bibnamefont {Fukuyama}},\ }\bibfield  {title}
  {\bibinfo {title} {{P}ossible quantum liquid crystal phases of helium
  monolayers},\ }\href {https://doi.org/10.1103/physrevb.94.180501} {\bibfield
  {journal} {\bibinfo  {journal} {Phys. Rev. B}\ }\textbf {\bibinfo {volume}
  {94}},\ \bibinfo {pages} {180501} (\bibinfo {year} {2016})}\BibitemShut
  {NoStop}%
\bibitem [{\citenamefont {Ny{\'{e}}ki}\ \emph {et~al.}(2017)\citenamefont
  {Ny{\'{e}}ki}, \citenamefont {Phillis}, \citenamefont {Ho}, \citenamefont
  {Lee}, \citenamefont {Coleman}, \citenamefont {Parpia}, \citenamefont
  {Cowan},\ and\ \citenamefont {Saunders}}]{Nyeki:2017ef}%
  \BibitemOpen
  \bibfield  {author} {\bibinfo {author} {\bibfnamefont {J.}~\bibnamefont
  {Ny{\'{e}}ki}}, \bibinfo {author} {\bibfnamefont {A.}~\bibnamefont
  {Phillis}}, \bibinfo {author} {\bibfnamefont {A.}~\bibnamefont {Ho}},
  \bibinfo {author} {\bibfnamefont {D.}~\bibnamefont {Lee}}, \bibinfo {author}
  {\bibfnamefont {P.}~\bibnamefont {Coleman}}, \bibinfo {author} {\bibfnamefont
  {J.}~\bibnamefont {Parpia}}, \bibinfo {author} {\bibfnamefont
  {B.}~\bibnamefont {Cowan}},\ and\ \bibinfo {author} {\bibfnamefont
  {J.}~\bibnamefont {Saunders}},\ }\bibfield  {title} {\bibinfo {title}
  {{I}ntertwined superfluid and density wave order in two-dimensional 4{H}e},\
  }\href {https://doi.org/10.1038/nphys4023} {\bibfield  {journal} {\bibinfo
  {journal} {Nat. Phys.}\ }\textbf {\bibinfo {volume} {13}},\ \bibinfo {pages}
  {455} (\bibinfo {year} {2017})}\BibitemShut {NoStop}%
\bibitem [{\citenamefont {Choi}\ \emph {et~al.}(2021)\citenamefont {Choi},
  \citenamefont {Zadorozhko}, \citenamefont {Choi},\ and\ \citenamefont
  {Kim}}]{Choi:2021vy}%
  \BibitemOpen
  \bibfield  {author} {\bibinfo {author} {\bibfnamefont {J.}~\bibnamefont
  {Choi}}, \bibinfo {author} {\bibfnamefont {A.~A.}\ \bibnamefont
  {Zadorozhko}}, \bibinfo {author} {\bibfnamefont {J.}~\bibnamefont {Choi}},\
  and\ \bibinfo {author} {\bibfnamefont {E.}~\bibnamefont {Kim}},\ }\bibfield
  {title} {\bibinfo {title} {{S}patially {M}odulated {S}uperfluid {S}tate in
  {T}wo-{D}imensional $^4${H}e {F}ilms},\ }\href
  {https://doi.org/10.1103/physrevlett.127.135301} {\bibfield  {journal}
  {\bibinfo  {journal} {Phys. Rev. Lett.}\ }\textbf {\bibinfo {volume} {127}},\
  \bibinfo {pages} {135301} (\bibinfo {year} {2021})}\BibitemShut {NoStop}%
\bibitem [{\citenamefont {Gordillo}\ and\ \citenamefont
  {Boronat}(2009)}]{Gordillo:2009jb}%
  \BibitemOpen
  \bibfield  {author} {\bibinfo {author} {\bibfnamefont {M.~C.}\ \bibnamefont
  {Gordillo}}\ and\ \bibinfo {author} {\bibfnamefont {J.}~\bibnamefont
  {Boronat}},\ }\bibfield  {title} {\bibinfo {title} {{$^4${H}e on a Single
  Graphene Sheet}},\ }\href {https://doi.org/10.1103/physrevlett.102.085303}
  {\bibfield  {journal} {\bibinfo  {journal} {Phys. Rev. Lett.}\ }\textbf
  {\bibinfo {volume} {102}},\ \bibinfo {pages} {085303} (\bibinfo {year}
  {2009})}\BibitemShut {NoStop}%
\bibitem [{\citenamefont {Gordillo}\ \emph {et~al.}(2011)\citenamefont
  {Gordillo}, \citenamefont {Cazorla},\ and\ \citenamefont
  {Boronat}}]{Gordillo:2011jb}%
  \BibitemOpen
  \bibfield  {author} {\bibinfo {author} {\bibfnamefont {M.~C.}\ \bibnamefont
  {Gordillo}}, \bibinfo {author} {\bibfnamefont {C.}~\bibnamefont {Cazorla}},\
  and\ \bibinfo {author} {\bibfnamefont {J.}~\bibnamefont {Boronat}},\
  }\bibfield  {title} {\bibinfo {title} {{Supersolidity in quantum films
  adsorbed on graphene and graphite}},\ }\href
  {http://journals.aps.org/prb/abstract/10.1103/PhysRevB.83.121406} {\bibfield
  {journal} {\bibinfo  {journal} {Phys. Rev. B}\ }\textbf {\bibinfo {volume}
  {83}},\ \bibinfo {pages} {121406(R)} (\bibinfo {year} {2011})}\BibitemShut
  {NoStop}%
\bibitem [{\citenamefont {Gordillo}\ and\ \citenamefont
  {Boronat}(2012)}]{Gordillo:2012fl}%
  \BibitemOpen
  \bibfield  {author} {\bibinfo {author} {\bibfnamefont {M.~C.}\ \bibnamefont
  {Gordillo}}\ and\ \bibinfo {author} {\bibfnamefont {J.}~\bibnamefont
  {Boronat}},\ }\bibfield  {title} {\bibinfo {title} {{Zero-temperature phase
  diagram of the second layer of $^4${H}e adsorbed on graphene}},\ }\href
  {https://doi.org/10.1103/physrevb.85.195457} {\bibfield  {journal} {\bibinfo
  {journal} {Phys. Rev. B}\ }\textbf {\bibinfo {volume} {85}},\ \bibinfo
  {pages} {195457} (\bibinfo {year} {2012})}\BibitemShut {NoStop}%
\bibitem [{\citenamefont {Kwon}\ and\ \citenamefont
  {Ceperley}(2012)}]{Kwon:2012ie}%
  \BibitemOpen
  \bibfield  {author} {\bibinfo {author} {\bibfnamefont {Y.}~\bibnamefont
  {Kwon}}\ and\ \bibinfo {author} {\bibfnamefont {D.~M.}\ \bibnamefont
  {Ceperley}},\ }\bibfield  {title} {\bibinfo {title} {{$^4${H}e adsorption on
  a single graphene sheet: Path-integral Monte Carlo study}},\ }\href
  {https://doi.org/10.1103/physrevb.85.224501} {\bibfield  {journal} {\bibinfo
  {journal} {Phys. Rev. B}\ }\textbf {\bibinfo {volume} {85}},\ \bibinfo
  {pages} {224501} (\bibinfo {year} {2012})}\BibitemShut {NoStop}%
\bibitem [{\citenamefont {Happacher}\ \emph {et~al.}(2013)\citenamefont
  {Happacher}, \citenamefont {Corboz}, \citenamefont {Boninsegni},\ and\
  \citenamefont {Pollet}}]{Happacher:2013ht}%
  \BibitemOpen
  \bibfield  {author} {\bibinfo {author} {\bibfnamefont {J.}~\bibnamefont
  {Happacher}}, \bibinfo {author} {\bibfnamefont {P.}~\bibnamefont {Corboz}},
  \bibinfo {author} {\bibfnamefont {M.}~\bibnamefont {Boninsegni}},\ and\
  \bibinfo {author} {\bibfnamefont {L.}~\bibnamefont {Pollet}},\ }\bibfield
  {title} {\bibinfo {title} {{Phase diagram of $^4${H}e on graphene}},\ }\href
  {https://doi.org/10.1103/physrevb.87.094514} {\bibfield  {journal} {\bibinfo
  {journal} {Phys. Rev. B}\ }\textbf {\bibinfo {volume} {87}},\ \bibinfo
  {pages} {094514} (\bibinfo {year} {2013})}\BibitemShut {NoStop}%
\bibitem [{\citenamefont {Gordillo}(2014)}]{Gordillo:2014cp}%
  \BibitemOpen
  \bibfield  {author} {\bibinfo {author} {\bibfnamefont {M.~C.}\ \bibnamefont
  {Gordillo}},\ }\bibfield  {title} {\bibinfo {title} {{Diffusion Monte Carlo
  calculation of the phase diagram of $^4${H}e on corrugated graphene}},\
  }\href {https://doi.org/10.1103/physrevb.89.155401} {\bibfield  {journal}
  {\bibinfo  {journal} {Phys. Rev. B}\ }\textbf {\bibinfo {volume} {89}},\
  \bibinfo {pages} {155401} (\bibinfo {year} {2014})}\BibitemShut {NoStop}%
\bibitem [{\citenamefont {{L. Vranje\v{s} Marki\'{c}}}\ \emph
  {et~al.}(2016)\citenamefont {{L. Vranje\v{s} Marki\'{c}}}, \citenamefont
  {Stipanovi\'{c}}, \citenamefont {Be\v{s}li\'{c}},\ and\ \citenamefont
  {Zillich}}]{Markic:2016dm}%
  \BibitemOpen
  \bibfield  {author} {\bibinfo {author} {\bibnamefont {{L. Vranje\v{s}
  Marki\'{c}}}}, \bibinfo {author} {\bibfnamefont {P.}~\bibnamefont
  {Stipanovi\'{c}}}, \bibinfo {author} {\bibfnamefont {I.}~\bibnamefont
  {Be\v{s}li\'{c}}},\ and\ \bibinfo {author} {\bibfnamefont {R.~E.}\
  \bibnamefont {Zillich}},\ }\bibfield  {title} {\bibinfo {title}
  {{Solidification of $^4${H}e clusters adsorbed on graphene}},\ }\href
  {https://doi.org/10.1103/physrevb.94.045428} {\bibfield  {journal} {\bibinfo
  {journal} {Phys. Rev. B}\ }\textbf {\bibinfo {volume} {94}},\ \bibinfo
  {pages} {045428} (\bibinfo {year} {2016})}\BibitemShut {NoStop}%
\bibitem [{\citenamefont {Huang}\ \emph {et~al.}(2009)\citenamefont {Huang},
  \citenamefont {Yan}, \citenamefont {Chen}, \citenamefont {Song},
  \citenamefont {Heinz},\ and\ \citenamefont {Hone}}]{Hone-PNAS}%
  \BibitemOpen
  \bibfield  {author} {\bibinfo {author} {\bibfnamefont {M.}~\bibnamefont
  {Huang}}, \bibinfo {author} {\bibfnamefont {H.}~\bibnamefont {Yan}}, \bibinfo
  {author} {\bibfnamefont {C.}~\bibnamefont {Chen}}, \bibinfo {author}
  {\bibfnamefont {D.}~\bibnamefont {Song}}, \bibinfo {author} {\bibfnamefont
  {T.~F.}\ \bibnamefont {Heinz}},\ and\ \bibinfo {author} {\bibfnamefont
  {J.}~\bibnamefont {Hone}},\ }\bibfield  {title} {\bibinfo {title} {{Phonon
  softening and crystallographic orientation of strained graphene studied by
  Raman spectroscopy}},\ }\href {https://doi.org/10.1073/pnas.0811754106}
  {\bibfield  {journal} {\bibinfo  {journal} {PNAS}\ }\textbf {\bibinfo
  {volume} {106}},\ \bibinfo {pages} {7304} (\bibinfo {year}
  {2009})}\BibitemShut {NoStop}%
\bibitem [{\citenamefont {Naumis}\ \emph {et~al.}(2017)\citenamefont {Naumis},
  \citenamefont {Barraza-Lopez}, \citenamefont {Oliva-Leyva},\ and\
  \citenamefont {Terrones}}]{Naumis2017}%
  \BibitemOpen
  \bibfield  {author} {\bibinfo {author} {\bibfnamefont {G.~G.}\ \bibnamefont
  {Naumis}}, \bibinfo {author} {\bibfnamefont {S.}~\bibnamefont
  {Barraza-Lopez}}, \bibinfo {author} {\bibfnamefont {M.}~\bibnamefont
  {Oliva-Leyva}},\ and\ \bibinfo {author} {\bibfnamefont {H.}~\bibnamefont
  {Terrones}},\ }\bibfield  {title} {\bibinfo {title} {Electronic and optical
  properties of strained graphene and other strained 2d materials: a review},\
  }\href {https://doi.org/10.1088/1361-6633/aa74ef} {\bibfield  {journal}
  {\bibinfo  {journal} {Rep. Prog. Phys.}\ }\textbf {\bibinfo {volume} {80}},\
  \bibinfo {pages} {096501} (\bibinfo {year} {2017})}\BibitemShut {NoStop}%
\bibitem [{\citenamefont {Zabel}\ \emph {et~al.}(2012)\citenamefont {Zabel},
  \citenamefont {Nair}, \citenamefont {Ott}, \citenamefont {Georgiou},
  \citenamefont {Geim}, \citenamefont {Novoselov},\ and\ \citenamefont
  {Casiraghi}}]{Zabel2012}%
  \BibitemOpen
  \bibfield  {author} {\bibinfo {author} {\bibfnamefont {J.}~\bibnamefont
  {Zabel}}, \bibinfo {author} {\bibfnamefont {R.~R.}\ \bibnamefont {Nair}},
  \bibinfo {author} {\bibfnamefont {A.}~\bibnamefont {Ott}}, \bibinfo {author}
  {\bibfnamefont {T.}~\bibnamefont {Georgiou}}, \bibinfo {author}
  {\bibfnamefont {A.~K.}\ \bibnamefont {Geim}}, \bibinfo {author}
  {\bibfnamefont {K.~S.}\ \bibnamefont {Novoselov}},\ and\ \bibinfo {author}
  {\bibfnamefont {C.}~\bibnamefont {Casiraghi}},\ }\bibfield  {title} {\bibinfo
  {title} {Raman spectroscopy of graphene and bilayer under biaxial strain:
  Bubbles and balloons},\ }\href {https://doi.org/10.1021/nl203359n} {\bibfield
   {journal} {\bibinfo  {journal} {Nano Letters}\ }\textbf {\bibinfo {volume}
  {12}},\ \bibinfo {pages} {617} (\bibinfo {year} {2012})}\BibitemShut
  {NoStop}%
\bibitem [{\citenamefont {Androulidakis}\ \emph {et~al.}(2015)\citenamefont
  {Androulidakis}, \citenamefont {Koukaras}, \citenamefont {Parthenios},
  \citenamefont {Kalosakas}, \citenamefont {Papagelis},\ and\ \citenamefont
  {Galiotis}}]{Androulidakis2015}%
  \BibitemOpen
  \bibfield  {author} {\bibinfo {author} {\bibfnamefont {C.}~\bibnamefont
  {Androulidakis}}, \bibinfo {author} {\bibfnamefont {E.~N.}\ \bibnamefont
  {Koukaras}}, \bibinfo {author} {\bibfnamefont {J.}~\bibnamefont
  {Parthenios}}, \bibinfo {author} {\bibfnamefont {G.}~\bibnamefont
  {Kalosakas}}, \bibinfo {author} {\bibfnamefont {K.}~\bibnamefont
  {Papagelis}},\ and\ \bibinfo {author} {\bibfnamefont {C.}~\bibnamefont
  {Galiotis}},\ }\bibfield  {title} {\bibinfo {title} {Graphene flakes under
  controlled biaxial deformation},\ }\bibfield  {journal} {\bibinfo  {journal}
  {Sci. Rep.}\ }\textbf {\bibinfo {volume} {5}},\ \href
  {https://doi.org/10.1038/srep18219} {10.1038/srep18219} (\bibinfo {year}
  {2015})\BibitemShut {NoStop}%
\bibitem [{\citenamefont {Przybytek}\ \emph {et~al.}(2010)\citenamefont
  {Przybytek}, \citenamefont {Cencek}, \citenamefont {Komasa}, \citenamefont
  {{\L}ach}, \citenamefont {Jeziorski},\ and\ \citenamefont
  {Szalewicz}}]{Przybytek:2010ol}%
  \BibitemOpen
  \bibfield  {author} {\bibinfo {author} {\bibfnamefont {M.}~\bibnamefont
  {Przybytek}}, \bibinfo {author} {\bibfnamefont {W.}~\bibnamefont {Cencek}},
  \bibinfo {author} {\bibfnamefont {J.}~\bibnamefont {Komasa}}, \bibinfo
  {author} {\bibfnamefont {G.}~\bibnamefont {{\L}ach}}, \bibinfo {author}
  {\bibfnamefont {B.}~\bibnamefont {Jeziorski}},\ and\ \bibinfo {author}
  {\bibfnamefont {K.}~\bibnamefont {Szalewicz}},\ }\bibfield  {title} {\bibinfo
  {title} {{R}elativistic and {Q}uantum {E}lectrodynamics {E}ffects in the
  {H}elium {P}air {P}otential},\ }\href
  {https://journals.aps.org/prl/abstract/10.1103/PhysRevLett.104.183003}
  {\bibfield  {journal} {\bibinfo  {journal} {Phys. Rev. Lett.}\ }\textbf
  {\bibinfo {volume} {104}},\ \bibinfo {pages} {183003} (\bibinfo {year}
  {2010})}\BibitemShut {NoStop}%
\bibitem [{\citenamefont {Nichols}\ \emph {et~al.}(2016)\citenamefont
  {Nichols}, \citenamefont {{Del Maestro}}, \citenamefont {Wexler},\ and\
  \citenamefont {Kotov}}]{Nichols:2016hd}%
  \BibitemOpen
  \bibfield  {author} {\bibinfo {author} {\bibfnamefont {N.~S.}\ \bibnamefont
  {Nichols}}, \bibinfo {author} {\bibfnamefont {A.}~\bibnamefont {{Del
  Maestro}}}, \bibinfo {author} {\bibfnamefont {C.}~\bibnamefont {Wexler}},\
  and\ \bibinfo {author} {\bibfnamefont {V.~N.}\ \bibnamefont {Kotov}},\
  }\bibfield  {title} {\bibinfo {title} {{Adsorption by design: Tuning
  atom-graphene van der Waals interactions via mechanical strain}},\ }\href
  {https://doi.org/10.1103/physrevb.93.205412} {\bibfield  {journal} {\bibinfo
  {journal} {Phys. Rev. B}\ }\textbf {\bibinfo {volume} {93}},\ \bibinfo
  {pages} {205412} (\bibinfo {year} {2016})}\BibitemShut {NoStop}%
\bibitem [{\citenamefont {Cencek}\ \emph {et~al.}(2012)\citenamefont {Cencek},
  \citenamefont {Przybytek}, \citenamefont {Komasa}, \citenamefont {Mehl},
  \citenamefont {Jeziorski},\ and\ \citenamefont {Szalewicz}}]{Cencek:2012iz}%
  \BibitemOpen
  \bibfield  {author} {\bibinfo {author} {\bibfnamefont {W.}~\bibnamefont
  {Cencek}}, \bibinfo {author} {\bibfnamefont {M.}~\bibnamefont {Przybytek}},
  \bibinfo {author} {\bibfnamefont {J.}~\bibnamefont {Komasa}}, \bibinfo
  {author} {\bibfnamefont {J.~B.}\ \bibnamefont {Mehl}}, \bibinfo {author}
  {\bibfnamefont {B.}~\bibnamefont {Jeziorski}},\ and\ \bibinfo {author}
  {\bibfnamefont {K.}~\bibnamefont {Szalewicz}},\ }\bibfield  {title} {\bibinfo
  {title} {{E}ffects of adiabatic, relativistic, and quantum electrodynamics
  interactions on the pair potential and thermophysical properties of helium},\
  }\href {https://doi.org/10.1063/1.4712218} {\bibfield  {journal} {\bibinfo
  {journal} {J. Chem. Phys.}\ }\textbf {\bibinfo {volume} {136}},\ \bibinfo
  {pages} {224303} (\bibinfo {year} {2012})}\BibitemShut {NoStop}%
\bibitem [{\citenamefont {Zimanyi}\ \emph {et~al.}(1994)\citenamefont
  {Zimanyi}, \citenamefont {Crowell}, \citenamefont {Scalettar},\ and\
  \citenamefont {Batrouni}}]{Zimanyi:1994rk}%
  \BibitemOpen
  \bibfield  {author} {\bibinfo {author} {\bibfnamefont {G.~T.}\ \bibnamefont
  {Zimanyi}}, \bibinfo {author} {\bibfnamefont {P.~A.}\ \bibnamefont
  {Crowell}}, \bibinfo {author} {\bibfnamefont {R.~T.}\ \bibnamefont
  {Scalettar}},\ and\ \bibinfo {author} {\bibfnamefont {G.~G.}\ \bibnamefont
  {Batrouni}},\ }\bibfield  {title} {\bibinfo {title} {{B}ose-{H}ubbard model
  and superfluid staircases in $^4${H}e films},\ }\href
  {https://doi.org/10.1103/physrevb.50.6515} {\bibfield  {journal} {\bibinfo
  {journal} {Phys. Rev. B}\ }\textbf {\bibinfo {volume} {50}},\ \bibinfo
  {pages} {6515} (\bibinfo {year} {1994})}\BibitemShut {NoStop}%
\bibitem [{\citenamefont {Murthy}\ \emph {et~al.}(1997)\citenamefont {Murthy},
  \citenamefont {Arovas},\ and\ \citenamefont {Auerbach}}]{Murthy-phase-diag}%
  \BibitemOpen
  \bibfield  {author} {\bibinfo {author} {\bibfnamefont {G.}~\bibnamefont
  {Murthy}}, \bibinfo {author} {\bibfnamefont {D.}~\bibnamefont {Arovas}},\
  and\ \bibinfo {author} {\bibfnamefont {A.}~\bibnamefont {Auerbach}},\
  }\bibfield  {title} {\bibinfo {title} {Superfluids and supersolids on
  frustrated two-dimensional lattices},\ }\href
  {https://doi.org/10.1103/PhysRevB.55.3104} {\bibfield  {journal} {\bibinfo
  {journal} {Phys. Rev. B}\ }\textbf {\bibinfo {volume} {55}},\ \bibinfo
  {pages} {3104} (\bibinfo {year} {1997})}\BibitemShut {NoStop}%
\bibitem [{\citenamefont {Lee}\ \emph {et~al.}(2008)\citenamefont {Lee},
  \citenamefont {Wei}, \citenamefont {Kysar},\ and\ \citenamefont
  {Hone}}]{Lee385}%
  \BibitemOpen
  \bibfield  {author} {\bibinfo {author} {\bibfnamefont {C.}~\bibnamefont
  {Lee}}, \bibinfo {author} {\bibfnamefont {X.}~\bibnamefont {Wei}}, \bibinfo
  {author} {\bibfnamefont {J.~W.}\ \bibnamefont {Kysar}},\ and\ \bibinfo
  {author} {\bibfnamefont {J.}~\bibnamefont {Hone}},\ }\bibfield  {title}
  {\bibinfo {title} {Measurement of the elastic properties and intrinsic
  strength of monolayer graphene},\ }\href
  {https://doi.org/10.1126/science.1157996} {\bibfield  {journal} {\bibinfo
  {journal} {Science}\ }\textbf {\bibinfo {volume} {321}},\ \bibinfo {pages}
  {385} (\bibinfo {year} {2008})}\BibitemShut {NoStop}%
\bibitem [{\citenamefont {Cao}\ \emph {et~al.}(2020)\citenamefont {Cao},
  \citenamefont {Feng}, \citenamefont {Han}, \citenamefont {Gao}, \citenamefont
  {Ly}, \citenamefont {Xu},\ and\ \citenamefont {Lu}}]{Cao2020}%
  \BibitemOpen
  \bibfield  {author} {\bibinfo {author} {\bibfnamefont {K.}~\bibnamefont
  {Cao}}, \bibinfo {author} {\bibfnamefont {S.}~\bibnamefont {Feng}}, \bibinfo
  {author} {\bibfnamefont {Y.}~\bibnamefont {Han}}, \bibinfo {author}
  {\bibfnamefont {L.}~\bibnamefont {Gao}}, \bibinfo {author} {\bibfnamefont
  {T.~H.}\ \bibnamefont {Ly}}, \bibinfo {author} {\bibfnamefont
  {Z.}~\bibnamefont {Xu}},\ and\ \bibinfo {author} {\bibfnamefont
  {Y.}~\bibnamefont {Lu}},\ }\bibfield  {title} {\bibinfo {title} {Elastic
  straining of free-standing monolayer graphene},\ }\href
  {https://doi.org/10.1038/s41467-019-14130-0} {\bibfield  {journal} {\bibinfo
  {journal} {Nat. Commun.}\ }\textbf {\bibinfo {volume} {11}},\ \bibinfo
  {pages} {284} (\bibinfo {year} {2020})}\BibitemShut {NoStop}%
\bibitem [{\citenamefont {Mohiuddin}\ \emph {et~al.}(2009)\citenamefont
  {Mohiuddin}, \citenamefont {Lombardo}, \citenamefont {Nair}, \citenamefont
  {Bonetti}, \citenamefont {Savini}, \citenamefont {Jalil}, \citenamefont
  {Bonini}, \citenamefont {Basko}, \citenamefont {Galiotis}, \citenamefont
  {Marzari}, \citenamefont {Novoselov}, \citenamefont {Geim},\ and\
  \citenamefont {Ferrari}}]{Geim-uniaxial}%
  \BibitemOpen
  \bibfield  {author} {\bibinfo {author} {\bibfnamefont {T.~M.~G.}\
  \bibnamefont {Mohiuddin}}, \bibinfo {author} {\bibfnamefont {A.}~\bibnamefont
  {Lombardo}}, \bibinfo {author} {\bibfnamefont {R.~R.}\ \bibnamefont {Nair}},
  \bibinfo {author} {\bibfnamefont {A.}~\bibnamefont {Bonetti}}, \bibinfo
  {author} {\bibfnamefont {G.}~\bibnamefont {Savini}}, \bibinfo {author}
  {\bibfnamefont {R.}~\bibnamefont {Jalil}}, \bibinfo {author} {\bibfnamefont
  {N.}~\bibnamefont {Bonini}}, \bibinfo {author} {\bibfnamefont {D.~M.}\
  \bibnamefont {Basko}}, \bibinfo {author} {\bibfnamefont {C.}~\bibnamefont
  {Galiotis}}, \bibinfo {author} {\bibfnamefont {N.}~\bibnamefont {Marzari}},
  \bibinfo {author} {\bibfnamefont {K.~S.}\ \bibnamefont {Novoselov}}, \bibinfo
  {author} {\bibfnamefont {A.~K.}\ \bibnamefont {Geim}},\ and\ \bibinfo
  {author} {\bibfnamefont {A.~C.}\ \bibnamefont {Ferrari}},\ }\bibfield
  {title} {\bibinfo {title} {Uniaxial strain in graphene by raman spectroscopy:
  $g$ peak splitting, {G}r\"uneisen parameters, and sample orientation},\
  }\href {https://doi.org/10.1103/PhysRevB.79.205433} {\bibfield  {journal}
  {\bibinfo  {journal} {Phys. Rev. B}\ }\textbf {\bibinfo {volume} {79}},\
  \bibinfo {pages} {205433} (\bibinfo {year} {2009})}\BibitemShut {NoStop}%
\bibitem [{\citenamefont {Amorim}\ \emph {et~al.}(2016)\citenamefont {Amorim},
  \citenamefont {Cortijo}, \citenamefont {de~Juan}, \citenamefont {Grushin},
  \citenamefont {Guinea}, \citenamefont {Guti{\'{e}}rrez-Rubio}, \citenamefont
  {Ochoa}, \citenamefont {Parente}, \citenamefont {Rold{\'{a}}n}, \citenamefont
  {San-Jose}, \citenamefont {Schiefele}, \citenamefont {Sturla},\ and\
  \citenamefont {Vozmediano}}]{maria}%
  \BibitemOpen
  \bibfield  {author} {\bibinfo {author} {\bibfnamefont {B.}~\bibnamefont
  {Amorim}}, \bibinfo {author} {\bibfnamefont {A.}~\bibnamefont {Cortijo}},
  \bibinfo {author} {\bibfnamefont {F.}~\bibnamefont {de~Juan}}, \bibinfo
  {author} {\bibfnamefont {A.~G.}\ \bibnamefont {Grushin}}, \bibinfo {author}
  {\bibfnamefont {F.}~\bibnamefont {Guinea}}, \bibinfo {author} {\bibfnamefont
  {A.}~\bibnamefont {Guti{\'{e}}rrez-Rubio}}, \bibinfo {author} {\bibfnamefont
  {H.}~\bibnamefont {Ochoa}}, \bibinfo {author} {\bibfnamefont
  {V.}~\bibnamefont {Parente}}, \bibinfo {author} {\bibfnamefont
  {R.}~\bibnamefont {Rold{\'{a}}n}}, \bibinfo {author} {\bibfnamefont
  {P.}~\bibnamefont {San-Jose}}, \bibinfo {author} {\bibfnamefont
  {J.}~\bibnamefont {Schiefele}}, \bibinfo {author} {\bibfnamefont
  {M.}~\bibnamefont {Sturla}},\ and\ \bibinfo {author} {\bibfnamefont
  {M.~A.~H.}\ \bibnamefont {Vozmediano}},\ }\bibfield  {title} {\bibinfo
  {title} {{Novel effects of strains in graphene and other two dimensional
  materials}},\ }\href
  {https://doi.org/http://dx.doi.org/10.1016/j.physrep.2015.12.006} {\bibfield
  {journal} {\bibinfo  {journal} {Phys. Rep.}\ }\textbf {\bibinfo {volume}
  {617}},\ \bibinfo {pages} {1} (\bibinfo {year} {2016})}\BibitemShut {NoStop}%
\bibitem [{\citenamefont {Pereira}\ \emph {et~al.}(2009)\citenamefont
  {Pereira}, \citenamefont {Neto},\ and\ \citenamefont
  {Peres}}]{10.1103/physrevb.80.045401}%
  \BibitemOpen
  \bibfield  {author} {\bibinfo {author} {\bibfnamefont {V.~M.}\ \bibnamefont
  {Pereira}}, \bibinfo {author} {\bibfnamefont {A.~H.~C.}\ \bibnamefont
  {Neto}},\ and\ \bibinfo {author} {\bibfnamefont {N.~M.~R.}\ \bibnamefont
  {Peres}},\ }\bibfield  {title} {\bibinfo {title} {{Tight-binding approach to
  uniaxial strain in graphene}},\ }\href
  {https://doi.org/10.1103/physrevb.80.045401} {\bibfield  {journal} {\bibinfo
  {journal} {Phys. Rev. B}\ }\textbf {\bibinfo {volume} {80}},\ \bibinfo
  {pages} {045401} (\bibinfo {year} {2009})}\BibitemShut {NoStop}%
\bibitem [{\citenamefont {Rold{\'{a}}n}\ \emph {et~al.}(2015)\citenamefont
  {Rold{\'{a}}n}, \citenamefont {Castellanos-Gomez}, \citenamefont
  {Cappelluti},\ and\ \citenamefont {Guinea}}]{Roldan_2015}%
  \BibitemOpen
  \bibfield  {author} {\bibinfo {author} {\bibfnamefont {R.}~\bibnamefont
  {Rold{\'{a}}n}}, \bibinfo {author} {\bibfnamefont {A.}~\bibnamefont
  {Castellanos-Gomez}}, \bibinfo {author} {\bibfnamefont {E.}~\bibnamefont
  {Cappelluti}},\ and\ \bibinfo {author} {\bibfnamefont {F.}~\bibnamefont
  {Guinea}},\ }\bibfield  {title} {\bibinfo {title} {Strain engineering in
  semiconducting two-dimensional crystals},\ }\href
  {https://doi.org/10.1088/0953-8984/27/31/313201} {\bibfield  {journal}
  {\bibinfo  {journal} {J. Phys. Condens. Mat.}\ }\textbf {\bibinfo {volume}
  {27}},\ \bibinfo {pages} {313201} (\bibinfo {year} {2015})}\BibitemShut
  {NoStop}%
\bibitem [{\citenamefont {Carrascoso}\ \emph {et~al.}(2022)\citenamefont
  {Carrascoso}, \citenamefont {Frisenda},\ and\ \citenamefont
  {Castellanos-Gomez}}]{Carrascoso2022}%
  \BibitemOpen
  \bibfield  {author} {\bibinfo {author} {\bibfnamefont {F.}~\bibnamefont
  {Carrascoso}}, \bibinfo {author} {\bibfnamefont {R.}~\bibnamefont
  {Frisenda}},\ and\ \bibinfo {author} {\bibfnamefont {A.}~\bibnamefont
  {Castellanos-Gomez}},\ }\bibfield  {title} {\bibinfo {title} {{Biaxial versus
  uniaxial strain tuning of single-layer MoS2}},\ }\href
  {https://doi.org/https://doi.org/10.1016/j.nanoms.2021.03.001} {\bibfield
  {journal} {\bibinfo  {journal} {Nano Mater. Sci.}\ }\textbf {\bibinfo
  {volume} {4}},\ \bibinfo {pages} {44} (\bibinfo {year} {2022})}\BibitemShut
  {NoStop}%
\bibitem [{\citenamefont {Meyer}\ \emph {et~al.}(2007)\citenamefont {Meyer},
  \citenamefont {Geim}, \citenamefont {Katsnelson}, \citenamefont {Novoselov},
  \citenamefont {Booth},\ and\ \citenamefont {Roth}}]{Meyer:2007ls}%
  \BibitemOpen
  \bibfield  {author} {\bibinfo {author} {\bibfnamefont {J.~C.}\ \bibnamefont
  {Meyer}}, \bibinfo {author} {\bibfnamefont {A.~K.}\ \bibnamefont {Geim}},
  \bibinfo {author} {\bibfnamefont {M.~I.}\ \bibnamefont {Katsnelson}},
  \bibinfo {author} {\bibfnamefont {K.~S.}\ \bibnamefont {Novoselov}}, \bibinfo
  {author} {\bibfnamefont {T.~J.}\ \bibnamefont {Booth}},\ and\ \bibinfo
  {author} {\bibfnamefont {S.}~\bibnamefont {Roth}},\ }\bibfield  {title}
  {\bibinfo {title} {{T}he structure of suspended graphene sheets},\ }\href
  {https://doi.org/10.1038/nature05545} {\bibfield  {journal} {\bibinfo
  {journal} {Nature}\ }\textbf {\bibinfo {volume} {446}},\ \bibinfo {pages}
  {60} (\bibinfo {year} {2007})}\BibitemShut {NoStop}%
\bibitem [{\citenamefont {Yamaguchi}\ \emph {et~al.}(2022)\citenamefont
  {Yamaguchi}, \citenamefont {Tajiri}, \citenamefont {Kumashita}, \citenamefont
  {Usami}, \citenamefont {Yamane}, \citenamefont {Sumiyama}, \citenamefont
  {Suzuki}, \citenamefont {Minoguchi}, \citenamefont {Sakurai},\ and\
  \citenamefont {Fukuyama}}]{Yamaguchi:2022qt}%
  \BibitemOpen
  \bibfield  {author} {\bibinfo {author} {\bibfnamefont {A.}~\bibnamefont
  {Yamaguchi}}, \bibinfo {author} {\bibfnamefont {H.}~\bibnamefont {Tajiri}},
  \bibinfo {author} {\bibfnamefont {A.}~\bibnamefont {Kumashita}}, \bibinfo
  {author} {\bibfnamefont {J.}~\bibnamefont {Usami}}, \bibinfo {author}
  {\bibfnamefont {Y.}~\bibnamefont {Yamane}}, \bibinfo {author} {\bibfnamefont
  {A.}~\bibnamefont {Sumiyama}}, \bibinfo {author} {\bibfnamefont
  {M.}~\bibnamefont {Suzuki}}, \bibinfo {author} {\bibfnamefont
  {T.}~\bibnamefont {Minoguchi}}, \bibinfo {author} {\bibfnamefont
  {Y.}~\bibnamefont {Sakurai}},\ and\ \bibinfo {author} {\bibfnamefont
  {H.}~\bibnamefont {Fukuyama}},\ }\bibfield  {title} {\bibinfo {title}
  {{S}tructural {S}tudy of {A}dsorbed {H}elium {F}ilms: {N}ew {A}pproach with
  {S}ynchrotron {R}adiation {X}-rays},\ }\href
  {https://doi.org/10.1007/s10909-021-02652-1} {\bibfield  {journal} {\bibinfo
  {journal} {J. Low Temp. Phys.}\ }\textbf {\bibinfo {volume} {208}},\ \bibinfo
  {pages} {441} (\bibinfo {year} {2022})}\BibitemShut {NoStop}%
\bibitem [{\citenamefont {Usami}\ \emph {et~al.}(2022)\citenamefont {Usami},
  \citenamefont {Toda}, \citenamefont {Nakamura}, \citenamefont {Matsui},\ and\
  \citenamefont {Fukuyama}}]{Usami:2022vs}%
  \BibitemOpen
  \bibfield  {author} {\bibinfo {author} {\bibfnamefont {J.}~\bibnamefont
  {Usami}}, \bibinfo {author} {\bibfnamefont {R.}~\bibnamefont {Toda}},
  \bibinfo {author} {\bibfnamefont {S.}~\bibnamefont {Nakamura}}, \bibinfo
  {author} {\bibfnamefont {T.}~\bibnamefont {Matsui}},\ and\ \bibinfo {author}
  {\bibfnamefont {H.}~\bibnamefont {Fukuyama}},\ }\bibfield  {title} {\bibinfo
  {title} {{A} simple {E}xperimental {S}etup for {S}imultaneous
  {S}uperfluid-{R}esponse and {H}eat-{C}apacity {M}easurements for {H}elium in
  {C}onfined {G}eometries},\ }\href
  {https://doi.org/10.1007/s10909-021-02658-9} {\bibfield  {journal} {\bibinfo
  {journal} {J. Low Temp. Phys.}\ }\textbf {\bibinfo {volume} {208}},\ \bibinfo
  {pages} {457} (\bibinfo {year} {2022})}\BibitemShut {NoStop}%
\bibitem [{\citenamefont {Ceperley}(1995)}]{Ceperley:1995gr}%
  \BibitemOpen
  \bibfield  {author} {\bibinfo {author} {\bibfnamefont {D.~M.}\ \bibnamefont
  {Ceperley}},\ }\bibfield  {title} {\bibinfo {title} {{Path integrals in the
  theory of condensed helium}},\ }\href
  {http://link.aps.org/doi/10.1103/RevModPhys.67.279} {\bibfield  {journal}
  {\bibinfo  {journal} {Rev. Mod. Phys.}\ }\textbf {\bibinfo {volume} {67}},\
  \bibinfo {pages} {279} (\bibinfo {year} {1995})}\BibitemShut {NoStop}%
\bibitem [{\citenamefont {Boninsegni}\ \emph {et~al.}(2006)\citenamefont
  {Boninsegni}, \citenamefont {{Prokof'ev}},\ and\ \citenamefont
  {Svistunov}}]{Boninsegni:2006ed}%
  \BibitemOpen
  \bibfield  {author} {\bibinfo {author} {\bibfnamefont {M.}~\bibnamefont
  {Boninsegni}}, \bibinfo {author} {\bibfnamefont {N.}~\bibnamefont
  {{Prokof'ev}}},\ and\ \bibinfo {author} {\bibfnamefont {B.}~\bibnamefont
  {Svistunov}},\ }\bibfield  {title} {\bibinfo {title} {{Worm Algorithm for
  Continuous-Space Path Integral Monte Carlo Simulations}},\ }\href
  {https://link.aps.org/doi/10.1103/PhysRevLett.96.070601} {\bibfield
  {journal} {\bibinfo  {journal} {Phys. Rev. Lett.}\ }\textbf {\bibinfo
  {volume} {96}},\ \bibinfo {pages} {070601} (\bibinfo {year}
  {2006})}\BibitemShut {NoStop}%
\bibitem [{\citenamefont {Nichols}\ \emph {et~al.}(2020)\citenamefont
  {Nichols}, \citenamefont {Prisk}, \citenamefont {Warren}, \citenamefont
  {Sokol},\ and\ \citenamefont {{Del Maestro}}}]{Nichols:2020of}%
  \BibitemOpen
  \bibfield  {author} {\bibinfo {author} {\bibfnamefont {N.~S.}\ \bibnamefont
  {Nichols}}, \bibinfo {author} {\bibfnamefont {T.~R.}\ \bibnamefont {Prisk}},
  \bibinfo {author} {\bibfnamefont {G.}~\bibnamefont {Warren}}, \bibinfo
  {author} {\bibfnamefont {P.}~\bibnamefont {Sokol}},\ and\ \bibinfo {author}
  {\bibfnamefont {A.}~\bibnamefont {{Del Maestro}}},\ }\bibfield  {title}
  {\bibinfo {title} {{D}imensional reduction of helium-4 inside argon-plated
  {M{C}M}-41 nanopores},\ }\href {https://doi.org/10.1103/physrevb.102.144505}
  {\bibfield  {journal} {\bibinfo  {journal} {Phys. Rev. B}\ }\textbf {\bibinfo
  {volume} {102}},\ \bibinfo {pages} {144505} (\bibinfo {year}
  {2020})}\BibitemShut {NoStop}%
\bibitem [{\citenamefont {{Del Maestro}}\ \emph {et~al.}(2022)\citenamefont
  {{Del Maestro}}, \citenamefont {Nichols}, \citenamefont {Prisk},
  \citenamefont {Warren},\ and\ \citenamefont {Sokol}}]{DelMaestro:2022wm}%
  \BibitemOpen
  \bibfield  {author} {\bibinfo {author} {\bibfnamefont {A.}~\bibnamefont {{Del
  Maestro}}}, \bibinfo {author} {\bibfnamefont {N.~S.}\ \bibnamefont
  {Nichols}}, \bibinfo {author} {\bibfnamefont {T.~R.}\ \bibnamefont {Prisk}},
  \bibinfo {author} {\bibfnamefont {G.}~\bibnamefont {Warren}},\ and\ \bibinfo
  {author} {\bibfnamefont {P.~E.}\ \bibnamefont {Sokol}},\ }\bibfield  {title}
  {\bibinfo {title} {{E}xperimental realization of one dimensional helium},\
  }\href {https://doi.org/10.1038/s41467-022-30752-3} {\bibfield  {journal}
  {\bibinfo  {journal} {Nat. Commun.}\ }\textbf {\bibinfo {volume} {13}},\
  \bibinfo {pages} {1038} (\bibinfo {year} {2022})}\BibitemShut {NoStop}%
\bibitem [{\citenamefont {Steele}(1973)}]{Steele:1973fo}%
  \BibitemOpen
  \bibfield  {author} {\bibinfo {author} {\bibfnamefont {W.~A.}\ \bibnamefont
  {Steele}},\ }\bibfield  {title} {\bibinfo {title} {The physical interaction
  of gases with crystalline solids},\ }\href
  {https://doi.org/10.1016/0039-6028(73)90264-1} {\bibfield  {journal}
  {\bibinfo  {journal} {Surf. Sci.}\ }\textbf {\bibinfo {volume} {36}},\
  \bibinfo {pages} {317} (\bibinfo {year} {1973})}\BibitemShut {NoStop}%
\bibitem [{\citenamefont {Nichols}(2021)}]{nichols_nathan_s_2021_6574043}%
  \BibitemOpen
  \bibfield  {author} {\bibinfo {author} {\bibfnamefont {N.~S.}\ \bibnamefont
  {Nichols}},\ }\bibfield  {title} {\bibinfo {title} {{3D lookup tables for
  helium-graphene interaction for isotropically strained graphene}},\
  }\bibfield  {journal} {\bibinfo  {journal} {Zenodo}\ }\href
  {https://doi.org/10.5281/zenodo.6574043} {10.5281/zenodo.6574043} (\bibinfo
  {year} {2021})\BibitemShut {NoStop}%
\bibitem [{\citenamefont {Fisher}\ \emph {et~al.}(1973)\citenamefont {Fisher},
  \citenamefont {Barber},\ and\ \citenamefont {Jasnow}}]{Fisher:1973zm}%
  \BibitemOpen
  \bibfield  {author} {\bibinfo {author} {\bibfnamefont {M.~E.}\ \bibnamefont
  {Fisher}}, \bibinfo {author} {\bibfnamefont {M.~N.}\ \bibnamefont {Barber}},\
  and\ \bibinfo {author} {\bibfnamefont {D.}~\bibnamefont {Jasnow}},\
  }\bibfield  {title} {\bibinfo {title} {{H}elicity {M}odulus, {S}uperfluidity,
  and {S}caling in {I}sotropic {S}ystems},\ }\href
  {https://doi.org/10.1103/physreva.8.1111} {\bibfield  {journal} {\bibinfo
  {journal} {Phys. Rev. A}\ }\textbf {\bibinfo {volume} {8}},\ \bibinfo {pages}
  {1111} (\bibinfo {year} {1973})}\BibitemShut {NoStop}%
\bibitem [{\citenamefont {Pollock}\ and\ \citenamefont
  {Ceperley}(1987)}]{Pollock:1987ta}%
  \BibitemOpen
  \bibfield  {author} {\bibinfo {author} {\bibfnamefont {E.}~\bibnamefont
  {Pollock}}\ and\ \bibinfo {author} {\bibfnamefont {D.~M.}\ \bibnamefont
  {Ceperley}},\ }\bibfield  {title} {\bibinfo {title} {{Path-Integral
  Computation of Superfluid Densities}},\ }\href
  {http://link.aps.org/doi/10.1103/PhysRevB.36.8343} {\bibfield  {journal}
  {\bibinfo  {journal} {Phys. Rev. B}\ }\textbf {\bibinfo {volume} {36}},\
  \bibinfo {pages} {8343} (\bibinfo {year} {1987})}\BibitemShut {NoStop}%
\bibitem [{\citenamefont {Prokof'ev}\ and\ \citenamefont
  {Svistunov}(2000)}]{Prokofev:2000ei}%
  \BibitemOpen
  \bibfield  {author} {\bibinfo {author} {\bibfnamefont {N.}~\bibnamefont
  {Prokof'ev}}\ and\ \bibinfo {author} {\bibfnamefont {B.}~\bibnamefont
  {Svistunov}},\ }\bibfield  {title} {\bibinfo {title} {{Two definitions of
  superfluid density}},\ }\href
  {http://link.aps.org/doi/10.1103/PhysRevB.61.11282} {\bibfield  {journal}
  {\bibinfo  {journal} {Phys. Rev. B}\ }\textbf {\bibinfo {volume} {61}},\
  \bibinfo {pages} {11282} (\bibinfo {year} {2000})}\BibitemShut {NoStop}%
\bibitem [{\citenamefont {Rousseau}(2014)}]{Rousseau:2014pv}%
  \BibitemOpen
  \bibfield  {author} {\bibinfo {author} {\bibfnamefont {V.~G.}\ \bibnamefont
  {Rousseau}},\ }\bibfield  {title} {\bibinfo {title} {{S}uperfluid density in
  continuous and discrete spaces: {A}voiding misconceptions},\ }\href
  {https://doi.org/10.1103/physrevb.90.134503} {\bibfield  {journal} {\bibinfo
  {journal} {Phys. Rev. B}\ }\textbf {\bibinfo {volume} {90}},\ \bibinfo
  {pages} {134503} (\bibinfo {year} {2014})}\BibitemShut {NoStop}%
\bibitem [{\citenamefont {Del~Maestro}(2022)}]{pimczm}%
  \BibitemOpen
  \bibfield  {author} {\bibinfo {author} {\bibfnamefont {A.}~\bibnamefont
  {Del~Maestro}},\ }\href {https://doi.org/10.5281/zenodo.7271913} {} (\bibinfo
  {year} {2022}),\ \bibinfo {note} {{Github Repository: Path Integral Quantum
  Monte Carlo \url{https://github.com/DelMaestroGroup/pimc}, Permanent link:
  \url{https://doi.org/10.5281/zenodo.7271913}}}\BibitemShut {NoStop}%
\bibitem [{\citenamefont {Young}(2012)}]{Young:2012ea}%
  \BibitemOpen
  \bibfield  {author} {\bibinfo {author} {\bibfnamefont {P.}~\bibnamefont
  {Young}},\ }\bibfield  {title} {\bibinfo {title} {Everything you wanted to
  know about data analysis and fitting but were afraid to ask},\ }\bibfield
  {journal} {\bibinfo  {journal} {arXiv:1210.3781}\ }\href
  {https://doi.org/10.48550/arxiv.1210.3781} {10.48550/arxiv.1210.3781}
  (\bibinfo {year} {2012})\BibitemShut {NoStop}%
\bibitem [{\citenamefont {Kim}\ and\ \citenamefont
  {Maestro}(2022{\natexlab{a}})}]{Zenodo:2022}%
  \BibitemOpen
  \bibfield  {author} {\bibinfo {author} {\bibfnamefont {S.~W.}\ \bibnamefont
  {Kim}}\ and\ \bibinfo {author} {\bibfnamefont {A.~D.}\ \bibnamefont
  {Maestro}},\ }\href {https://doi.org/10.5281/zenodo.7271852} {\bibinfo
  {title} {{QMC Raw Data for Superfulid Helium Adsorbed on Strained Graphene}}}
  (\bibinfo {year} {2022}{\natexlab{a}})\BibitemShut {NoStop}%
\bibitem [{\citenamefont {Kim}\ and\ \citenamefont
  {Maestro}(2022{\natexlab{b}})}]{sang_wook_kim_2022_7294692}%
  \BibitemOpen
  \bibfield  {author} {\bibinfo {author} {\bibfnamefont {S.~W.}\ \bibnamefont
  {Kim}}\ and\ \bibinfo {author} {\bibfnamefont {A.~D.}\ \bibnamefont
  {Maestro}},\ }\href {https://doi.org/10.5281/zenodo.7294692} {\bibinfo
  {title} {Github repository:
  \url{https://github.com/DelMaestroGroup/papers-code-Superfluid4HeStrainGraphene}}}
  (\bibinfo {year} {2022}{\natexlab{b}})\BibitemShut {NoStop}%
\end{thebibliography}%





\ifarXiv
    

\section*{Supplementary material}

# Supplementary Information for "Strain-induced superfluid transition for atoms on graphene"

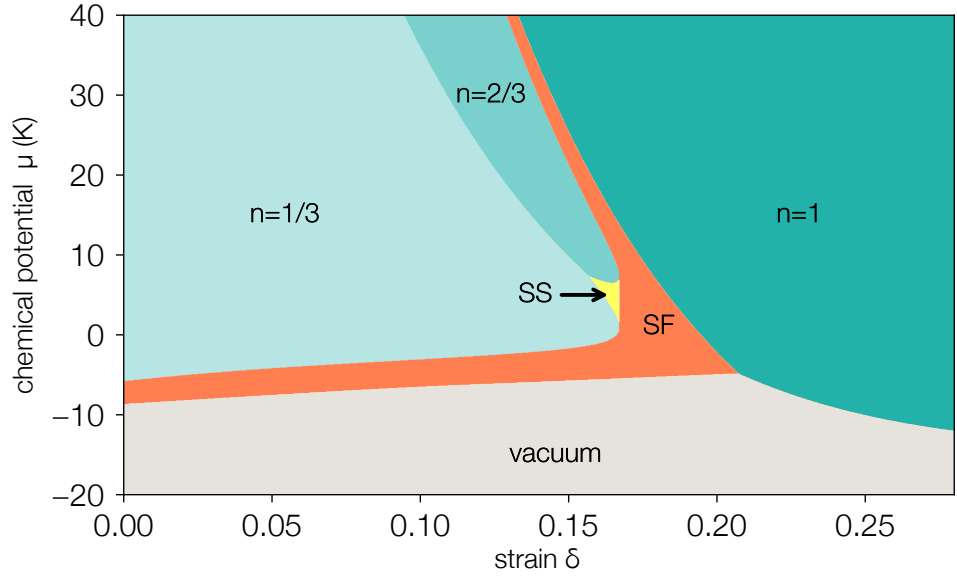

FIG. S1. **Mean field phase diagram of the extended 2D Bose-Hubbard model on the triangular lattice presented in physical units.** The dimensionless phase diagram in the main text (Figure 1**a**) can be converted to chemical potential $\mu$ and strain $\delta$ using the strain dependence of interactions $V, V'$ and hopping $t$ computed via Hartree–Fock. The phase diagram which includes commensurate solid phases at filling fractions $n = 1/3, 2/3$, and $1$ as well as superfluid (SF) and supersolid (SS) phases bears remarkable similarity to that computed via quantum Monte Carlo (Figure 3).

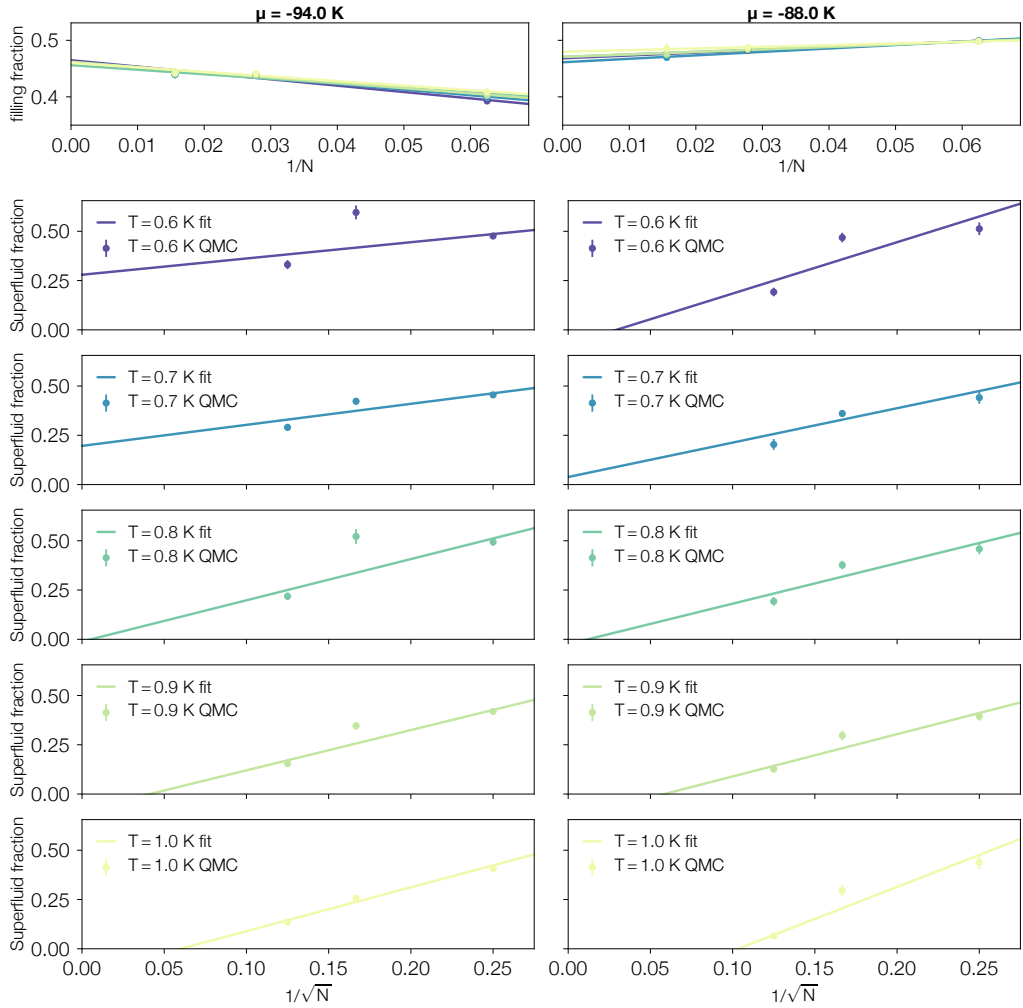

FIG. S2. **Finite size scaling of the particle and superfluid density.** For each $(\delta, \mu)$ data point, the existence of a superfluid phase in the thermodynamic limit is investigated by finite size scaling of Quantum Monte Carlo (QMC) data points. Columns show this procedure at two different values of $\mu$ for $\delta = 0.12$. For $\mu = -94.0$ K, superfluidity persists below $T = 0.7$ K, while no region of superfluidity if found for $\mu = -88$ K in the thermodynamic limit. The form of the linear scaling functions (lines) for the filling fraction (top row) and superfluid fraction $\rho_s/\rho$ (rows 2–6) is described in the Methods section of the main text. Error bars represent standard statistical uncertainties in Quantum Monte Carlo.

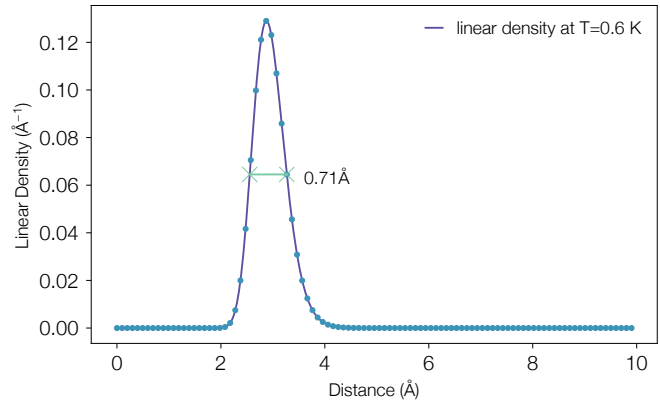

FIG. S3. **Effective dimensionality of the adsorbed $^4$He layer.** The density of $^4$He as a function of the distance above the graphene membrane at $\delta = 0.15$ for $\mu/k_B = -93$ K at $T = 0.6$ K. The full width half maximum of 0.71 Å of the adsorbed layer strongly supports the effective low energy treatment within the 2D Bose–Hubbard picture.

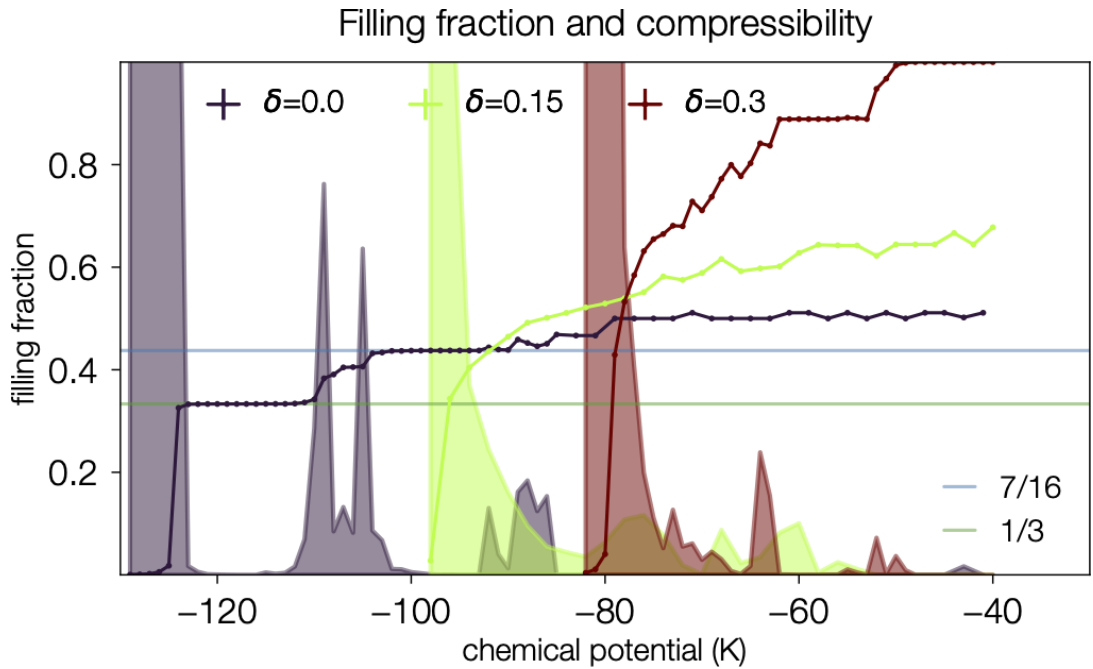

FIG. S4. **Adsorption transition for $^4$He on strained graphene.** The filling fraction $n$ as a function of chemical potential $\mu$ for three values of strain at $T = 1.0\,\text{K}$ (symbols). Adsorption curves are obtained from simulations at $N_\text{G} = 48, 64, 90$. The filled region in the background corresponds to the normalized compressibility $\kappa \propto \langle (N - \langle N \rangle)^2 \rangle$. Plateaus in the adsorption curves correspond to incompressible solid phases which may exist at commensurate filling fractions (indicated by horizontal lines at $n = 1/3$ and $n = 7/16$). A state with unit filling ($n = 1$) is only possible at extreme strain where the distance between triangular lattice adsorption sites has been stretched to a regime where the interactions between helium atoms become attractive. For $\delta = 0.15$ the qualitatively different onset behavior (characterized by the continuous increase in density and high compressibility) corresponds to the superfluid region of the phase diagram described in the main text Figure 3. Error bars represent statistical uncertainties from Quantum Monte Carlo sampling.


\fi

\end{document}